\documentclass[5p,twocolumn,times,number,dvipdfmx]{elsarticle}
\newcommand{\fref}[1]{Fig.~\ref{#1}}
\newcommand{\Fref}[1]{Figure~\ref{#1}}
\newcommand{\frefs}[2]{Figs.~\ref{#1}--\subref{#2}}
\newcommand{\Frefs}[2]{Figures~\ref{#1}--\subref{#2}}

\newcommand{\Frefsa}[2]{Figures~\ref{#1} and \ref{#2}}

\newcommand{\Frefss}[2]{Figures~\ref{#1}--\ref{#2}}
\usepackage{graphicx}
\usepackage{comment}
\usepackage{amsmath}   
\usepackage{lineno,hyperref,subfiles,multicol,subcaption}
\usepackage{color}

\modulolinenumbers[5]
\journal{NIM-A}
\hypersetup{
colorlinks=true,
linkcolor=blue,
citecolor=blue,
urlcolor=blue,
}
\begin{document}

\begin{frontmatter}
\title{Timing resolution of a plastic scintillator counter read out by radiation damaged 
SiPMs connected in series}

\author[add2,add3]{G.~Boca}
\author[add2]{P.~W.~Cattaneo}
\author[add4]{M.~De~Gerone}
\author[add4,add5]{F.~Gatti}
\author[add8]{M.~Nakao}
\author[add8]{M.~Nishimura}
\author[add9]{W.~Ootani}
\author[add2]{M.~Rossella}
\author[add9]{Y.~Uchiyama}
\author[add8]{M.~Usami\corref{cor}}
\ead{usami@icepp.s.u-tokyo.ac.jp}
\author[add8]{K.~Yanai}

\cortext[cor]{Corresponding author}

\address[add2]{INFN Sezione di Pavia, Via A. Bassi 6, 27100 Pavia, Italy}
\address[add3]{Dipartimento di Fisica, Universit\`a degli Studi di Pavia, Via A. Bassi 6, 27100 Pavia, Italy}
\address[add4]{INFN Sezione di Genova, Via Dodecaneso 33, 16146 Genova, Italy}
\address[add5]{Dipartimento di Fisica, Universit\`a degli Studi di Genova, Via Dodecaneso 33, 16146 Genova, Italy}
\address[add8]{Department of Physics, The University of Tokyo, 7-3-1 Hongo, Bunkyo-ku, Tokyo 113-0033, Japan}
\address[add9]{ICEPP, The University of Tokyo, 7-3-1 Hongo, Bunkyo-ku, Tokyo 113-0033, Japan}

\begin{abstract}
 
This paper discusses the effects of radiation damage to SiPMs on the performances of plastic scintillator 
counters with series-connected SiPM readout, focusing on timing measurements. 
The performances of a counter composed of a $120 \times 40 \times5~\mathrm{mm}^3$ scintillator tile read 
out by two sets of six SiPMs from AdvanSiD connected in series attached on the short sides are presented,
for different combinations of SiPMs at various levels of irradiation. 
Firstly, six SiPMs were equally irradiated with electrons from $^{90}$Sr 
sources up to a fluence of $\Phi_\mathrm{e^-}\approx 3 \times 10^{12}~\mathrm{cm}^{-2}$.
The timing resolution of the counter gradually deteriorated by the increase in dark current. 
The dark current and the deterioration were reduced 
when the counter was cooled from 30$^\circ$C to 10$^\circ$C.
Secondly, 33 SiPMs were irradiated with reactor neutrons. The characteristics of 
counters read out by series-connected SiPMs with non-uniform damage levels, 
were investigated. 
The signal pulse height, the time response, and the timing resolution 
depend on the hit position in the counter, when SiPMs' irradiation is not uniform.
\end{abstract}

\end{frontmatter}
\tableofcontents


\section{Introduction}

Scintillator detectors read out by Silicon PhotoMultipliers (SiPMs) are used in a 
wide range of experiments to detect charged particles in particular when high timing
resolution is required.
Series connection of SiPMs is found to be effective to achieve high 
timing resolutions \cite{Series_Connection_Study_MEG},
because it reduces the SiPMs' total capacitance and therefore
reduces the rise time of the output signal.
Examples are the pixelated Timing Counter (pTC) in the MEG II 
experiment \cite{MEG_II_Experiment}, 
the electromagnetic calorimeter in the Mu2e experiment \cite{Mu2e_Experiment},
and the TOF detector in the PANDA experiment \cite{PANDA_TOF_Detector}.

Many studies on irradiation damage on SiPMs are available (recent studies 
are summarised in \cite{Radiation_Damage_Recent_Paper}). Nevertheless, a detailed study 
on its effect on the timing resolution of scintillator counters read out by series-connected 
SiPMs has not been presented yet.
In this work, the effect of radiation damage to SiPMs on the performance of a counter consisting of a 
scintillator tile 
read out by two sets of six series-connected SiPMs on opposite sides is investigated, focusing on the timing resolution. 
First, the effect of the increase in dark current is investigated with six equally-irradiated SiPMs.
Next, additional impact on timing measurements due to non-uniformity of irradiation is
investigated with several combinations of differently-irradiated SiPMs.
In an experimental setup non-uniform damage among series-connected SiPMs can occur if the hit rate 
in the counter depends on the position of the SiPMs. 

\section{Radiation damage to SiPMs and series connection}

\subsection{Radiation damage to silicon sensors}

Collisions of energetic particles with silicon bulk cause displacement of  the atoms, i.e.\ lattice defects.
This bulk damage can be classified into two types: point defects and cluster defects. Their relative weight depends on the type and the energy of incident particles.
Low energy electrons, such as those from $^{90}$Sr sources (maximum energy at 2.28~MeV 
from cascade $^{90}$Y decay), mainly produce point defects while hadrons such as 1-MeV neutrons produce cluster defects.
More energetic electrons with $O(10~\mathrm{MeV})$ energy produce both types with the relative weight of cluster defects increasing with the energy.

One of the major effects of the bulk damage to the sensor characteristics is an increase in bulk dark current.
Therefore, the damage level is often quantified using the dark current increase.
In case of SiPMs, the bulk dark current undergoes multiplication and results in dark counts, which are indistinguishable from light-induced counts, that may degrade timing measurement.

\subsection{Scintillator counters with series-connected SiPMs readout}
 
The series connection of multiple SiPMs shows some peculiar features.
When $n$ SiPMs are connected in series, the total sensor capacitance reduces by 
a factor of $n$, while the effective charge gain of each SiPM also reduces by the same factor. 
The voltage needed to apply the same bias voltage $V_\mathrm{bias}$ to each SiPM 
becomes $n$ times higher.
A common current flows through all the SiPMs and balances the bias voltages among the SiPMs. 
As in the case of parallel connection, not only the photon statistics but also the dark counts are summed.
 
Sect.~\ref{Sec:MEGIIpTC} presents the pTC in the MEG II experiment because
the design of our test counter is based on the design of a pTC counter.

\subsubsection{MEG II pTC}
\label{Sec:MEGIIpTC}
The MEG II pTC, dedicated to measuring the timing of positrons from muon decays, is composed of 512 plastic scintillator counters. A single counter is a two-side readout plastic scintillator (BC422) with dimensions of $120 \times 40 (50) \times 5~\mathrm{mm^3}$; each side is read out by six series-connected SiPMs with cell size $50\times 50~\mathrm{\muup m^2}$ and $3\times3~\mathrm{mm^2}$ active area\footnote{ASD-NUV3S-P High-Gain from AdvanSiD.}.
Each side detects roughly 50--100 photoelectrons, depending on the scintillator light yield and the photo-detection efficiency (PDE) of the SiPMs, achieving 70--90 ps timing resolution with the two-side measurement.

The pTC was operated in pilot runs of the MEG II experiment for a month using the nominal intensity positive muon beam ($7\times10^7~\mathrm{s}^{-1}$)~\cite{pilotrun2016}.
During the beam time, an increase in the SiPMs' dark currents was observed.
By linearly extrapolating the increase to the planned three-year data-taking period, the dark currents, a few $\mathrm{\muup A}$ before the exposure, are expected to reach $I_\mathrm{dark}\sim 100~\mathrm{\muup A}$ on average at the operating bias voltages 
at 30$^\circ$C.    
The positron flux was measured by the counters. It depends on the position of the counters; the counters at the highest hit-rate region ($\sim100$ kHz) will be exposed to $\Phi_\mathrm{e^+} \sim 8 \times 10^{10}~\mathrm{cm^{-2}}$ of $\sim 50$~MeV positrons with an absorbed dose of $\sim 20$~Gy in the three-year run. The total fluence roughly corresponds  to the 1-MeV neutron equivalent fluence $\Phi_\mathrm{eq} \approx 4\times 10^{9}~\mathrm{cm^{-2}}$, converted by using an ``effective NIEL''  calculated using molecular dynamics simulations~\cite{EffectiveNIEL}.
The fluence is not uniform inside a counter. It depends on the distance from the beam axis, and the inner-most SiPM will be exposed to roughly triple the flux compared to that of the outer-most SiPM.
Preliminary results on the effects of radiation damage were reported in \cite{Pisa_Proceedings}.

\section{Experimental setup}

\subsection{Samples and irradiation}
SiPM samples (ASD-NUV3S-P High-Gain) were irradiated with electrons and neutrons at different fluences.
ASD-NUV3S-P High-Gain is based on a p-on-n structure for detection of near ultraviolet light and the properties are summarised in Table~\ref{tab:SiPM}.

\begin{table}[!t]
\caption{Properties of ASD-NUV3S-P High-Gain.}
\label{tab:SiPM}
\centering
  \begin{minipage}{0.8\linewidth}
   \renewcommand{\thefootnote}{\alph{footnote})}	
   \renewcommand{\thempfootnote}{\alph{mpfootnote})}	
\centering
\small
\begin{tabular}{@{}lr}
\hline
Parameter  & Value \\
\hline \hline
Active area\footnotemark[1] ($\mathrm{mm}^2$)         & $3 \times 3$ \\
Cell size\footnotemark[1] ($\muup\mathrm{m}^2$)  & $50 \times 50$ \\
Peak sensitivity wavelength\footnotemark[1] (nm)          & $420$ \\
Breakdown voltage\footnotemark[1] (V)                         & $24 \pm 0.3$ \\
Gain\footnotemark[1]  (at $V_\mathrm{over} = 3$ V)     & $3.3 \times 10^6$ \\
Quench resistance\footnotemark[2] (M$\Omega$)          & $1.1$ \\
Recharge time constant\footnotemark[2] (ns)                  & $124$ \\                 
\hline
\end{tabular}
  \footnotetext[1]{From AdvanSiD data-sheet.}
  \footnotetext[2]{Measured value~\cite{VCI2016}.}
\end{minipage}
\end{table}

\subsubsection*{Electron-irradiated samples}
Six SiPMs (\#1--\#6) were irradiated to the same fluence using two 37-MBq $\mathrm{^{90}Sr}$ sources. 
The irradiation was separated into four steps, each 70 hours long.  The two sources were swapped alternately between each step to irradiate uniformly all the samples. 
During the irradiation, no bias voltage was applied to the SiPMs.
Between the steps, the measurement described in Sect.~\ref{sec:measurements} was performed. 
The total fluence in the 280-hour exposure was measured to be $\Phi_{\mathrm{e^-}}\approx 3 \times 10^{12}~\mathrm{cm^{-2}}$ with a total dose of 1~kGy,\footnote{The total dose was evaluated with a simulation based on Geant4.} equivalent to $\Phi_\mathrm{eq} \approx 3\times 10^{9}~\mathrm{cm^{-2}}$ using the effective NIEL value~\cite{EffectiveNIEL}.

\subsubsection*{Neutron-irradiated samples}

A set of 33 SiPMs was irradiated with neutrons with kinetic energies ranging from 0.5~MeV to 16~MeV using the reactor neutron facilities at the Laboratory of
Applied Nuclear Energy (LENA) of the University of Pavia. The set was divided into sub-groups for different fluence levels ranging from $\Phi_\mathrm{eq} \approx 8.7 \times 10^8$ to  $\Phi_\mathrm{eq}\approx 5.5 \times 10^{13}~\mathrm{cm^{-2}}$.
The 1-MeV neutron equivalent fluences were calculated from the differential neutron flux of the reactor following \cite{ASTM-E-722-93}.
Among them, two samples with $\Phi_\mathrm{eq}\approx 8.7 \times 10^8~\mathrm{cm^{-2}}$ (\#7, \#8) and two with $\Phi_\mathrm{eq}\approx 5.5 \times 10^9~\mathrm{cm^{-2}}$ (\#9, \#10) were used in this study.

\begin{figure}[tb]
		\centering
		\includegraphics[clip, width=1.\columnwidth, trim=400 300 400 300]{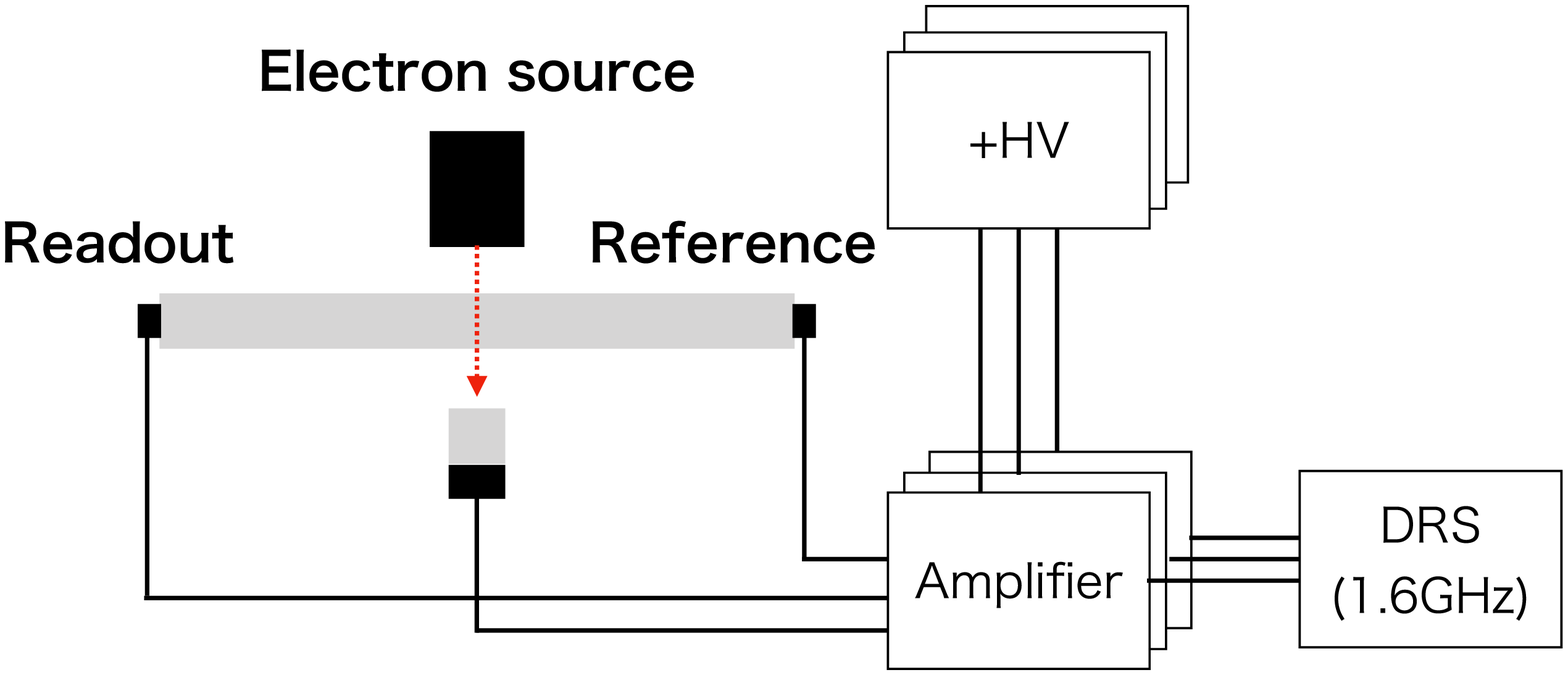}
		\caption{Setup for timing resolution measurements with a scintillator tile read out 
             by the six series-connected SiPMs under test. Six non-irradiated SiPMs were located on the opposite 
             side of the tile as reference.}
             \label{Fig:MeasurementSetup}
\end{figure}
\begin{figure}[tb]
		\centering
		\includegraphics[clip, width=1.\columnwidth]{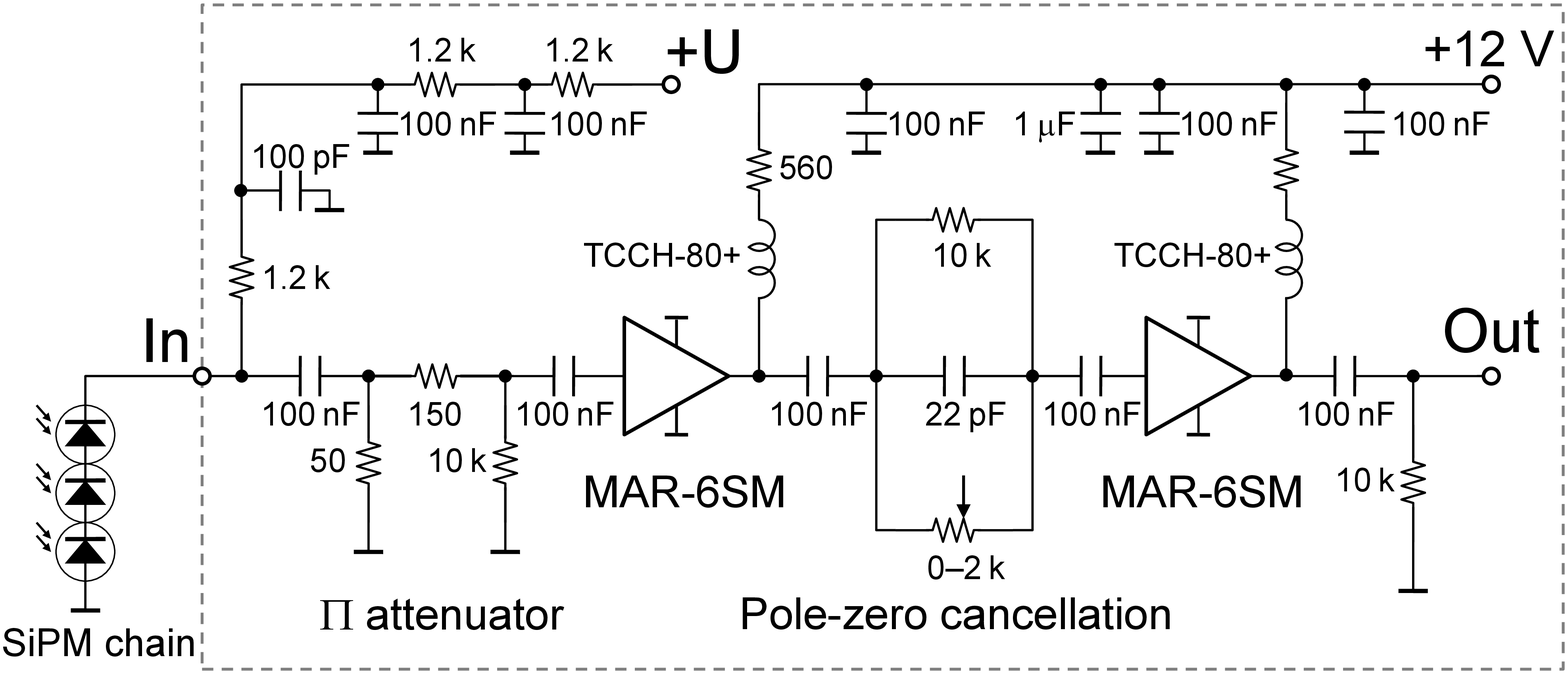}
		\caption{Schematics of the amplifier. The pole-zero cancellation filter can be applied by tuning the adjustable resistor in the circuit~\cite{Series_Connection_Study_MEG}.}\label{Fig:PSI_amp}
\end{figure}
\begin{figure}[tb]
		\centering
		\includegraphics[clip, width=0.99\columnwidth]{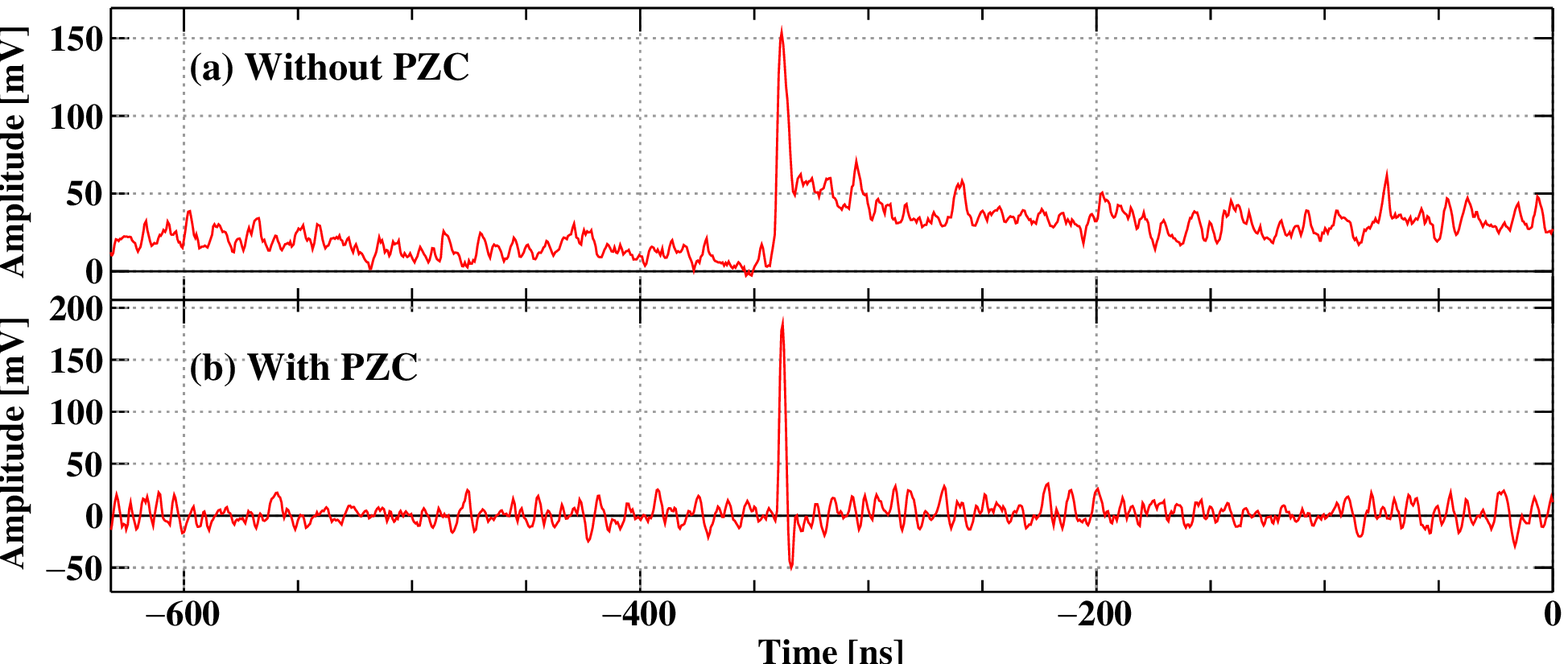}
		\caption{Comparison of typical waveforms with and without the pole-zero cancellation filter. They are taken with irradiated SiPMs (pattern C in Sect.~\ref{sec:differently-damaged-measurement}, see the text for the details).
		The pulse is attenuated by a factor of two in (a). The long tail in (a) is due to the recharge current through the quench resistance.}
              \label{Fig:Waveform_Example_PZ}
 \end{figure}
 
\subsection{Timing measurements} \label{sec:measurements}

The main topic of this paper is the measurement of the timing resolution of a counter 
equipped with irradiated SiPMs.
\Fref{Fig:MeasurementSetup} shows the measurement setup. 
The counter consists of a $40 \times 120 \times 5 ~\mathrm{mm^3}$ scintillator tile (BC422), whose
scintillation photons are read out by six series-connected SiPMs on each short side:
on one side, the SiPMs under test and, on the opposite one, non-irradiated 
SiPMs as reference.
The tile and the SiPMs are coupled with optical grease and are aligned through a custom-made 
jig produced with a 3D printer. 
The electrical contacts with the SiPMs\footnote{They are surface-mount type.} are made by means of spring probe pins 
so that the series circuit can be made each time without soldering.

A 3.7-MBq $\mathrm{^{90}Sr}$ source emitted electrons into the counter. 
The trigger signals were generated by a $5 \times 5 \times 5 ~\mathrm{mm^3}$ 
scintillator (BC422) trigger counter read out by one SiPM (Hamamatsu Photonics S10362-33-050C),
located under the scintillator. 
A fraction of the electrons is energetic enough to reach the trigger counter and release 
energy sufficient to fire the trigger.

The trigger counter has a much better timing resolution 
($\sigma_{t_\mathrm{trigger}}\sim 30$ ps) than the test counter~\cite{Series_Connection_Study_MEG},
and therefore, its timing measurement ($t_\mathrm{trigger}$) can be used
as a reference of the electron hit timing on the test counter ($t_\mathrm{signal}$).
The time response of the six series-connected SiPMs under test was characterised by the mean value of  
$t_\mathrm{signal} - t_\mathrm{trigger}$ distribution, $\mu_\mathrm{t}$ or ``time centre'', and the 
timing resolution from its standard deviation $\sigma_t$.\footnote{The counter resolution 
becomes $\sqrt{2}$ times better with the two-side readout.}

The output signals were amplified and shaped through a two-staged voltage amplifier with 
a pole-zero cancellation (PZC) filter, shown in \fref{Fig:PSI_amp}.
\Fref{Fig:Waveform_Example_PZ} shows typical waveforms after irradiation with and without PZC. 
The baseline is stabilised by the filter and the dark count effect on the timing measurement is significantly suppressed.
The PZC is effective for a good timing measurement especially under high dark count rate condition.
In this study, the PZC was always active.

The signal waveforms were recorded using the Domino Ring Sampler waveform digitiser (DRS4)~\cite{DRS}, at a sampling speed of 1.6 GS/s.              
The pulse timing was measured through the digital constant fraction discriminator method (CFD) with a cubic interpolation between the sampling points. 
              
Some parameters of SiPMs 
have  significant temperature dependence. 
Therefore, the measurements were performed at 30$^\circ$C and 10$^\circ$C to evaluate 
the effect of the operating temperature.

\section{Dark current measurement and timing resolution}
\label{sec:equally-damaged-measurement}

\subsection{I-V curves}

\begin{figure*}[ptb]
	\begin{minipage}[t]{\columnwidth}
		\centering
		\includegraphics[clip, width=0.9\columnwidth]{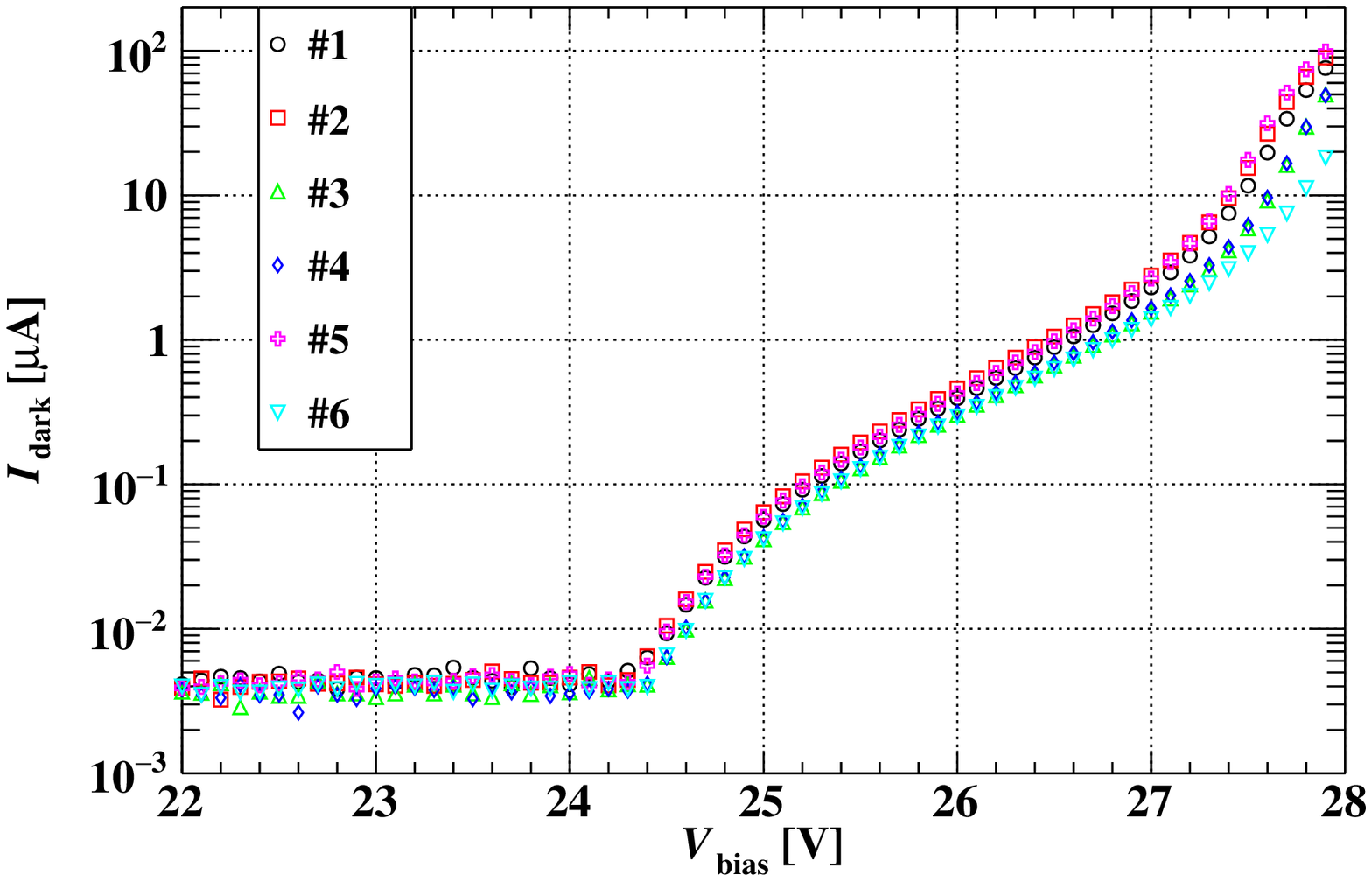}
		\subcaption{Before irradiation.}\label{Fig:0hour}
	\end{minipage}
		\hfil
	\begin{minipage}[t]{\columnwidth}
		\centering
		\includegraphics[clip, width=0.9\columnwidth]{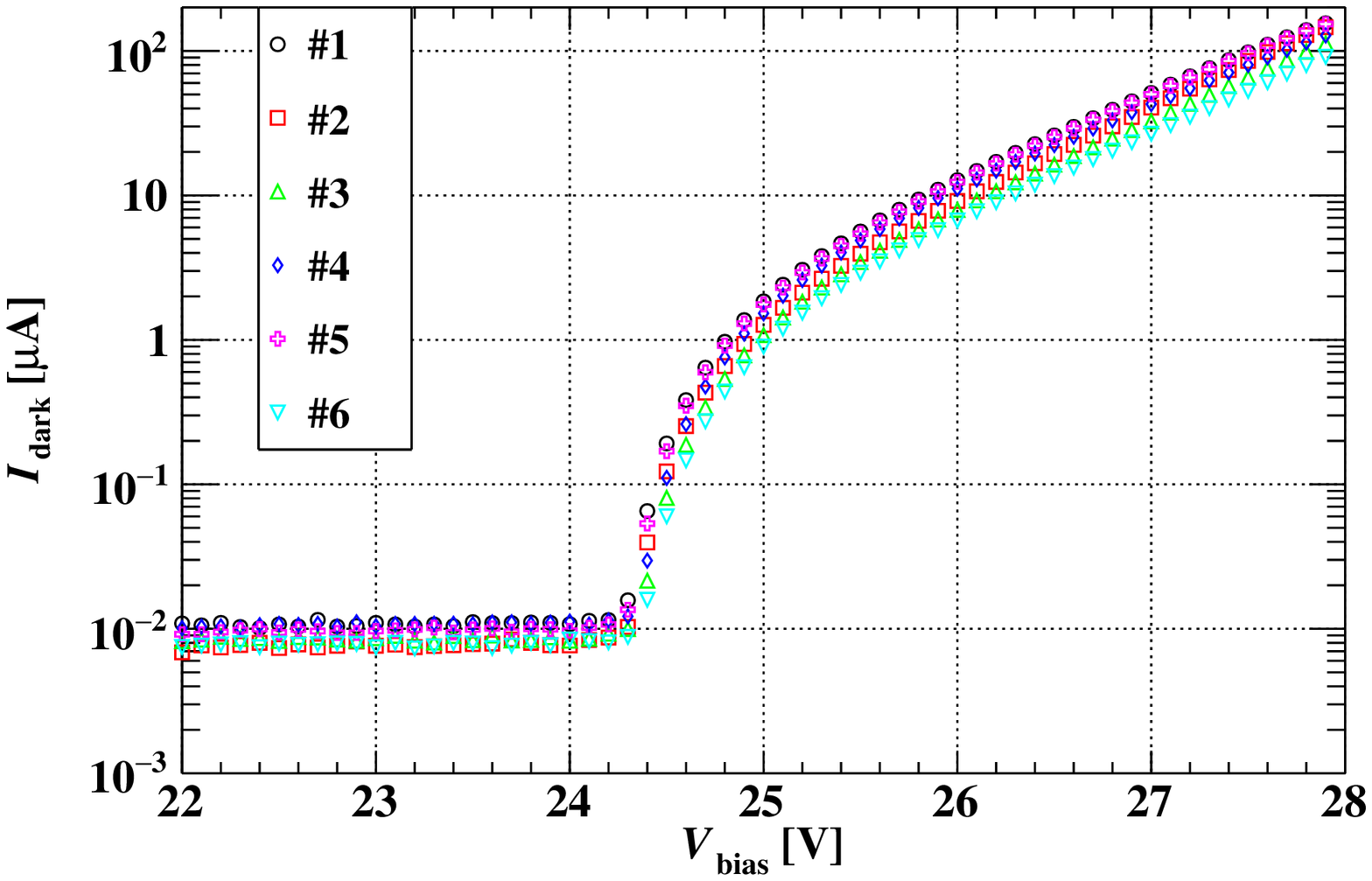}
		\subcaption{After 70 hours irradiation.}
	\end{minipage}
	
	\begin{minipage}[t]{\columnwidth}
		\centering
		\includegraphics[clip, width=0.9\columnwidth]{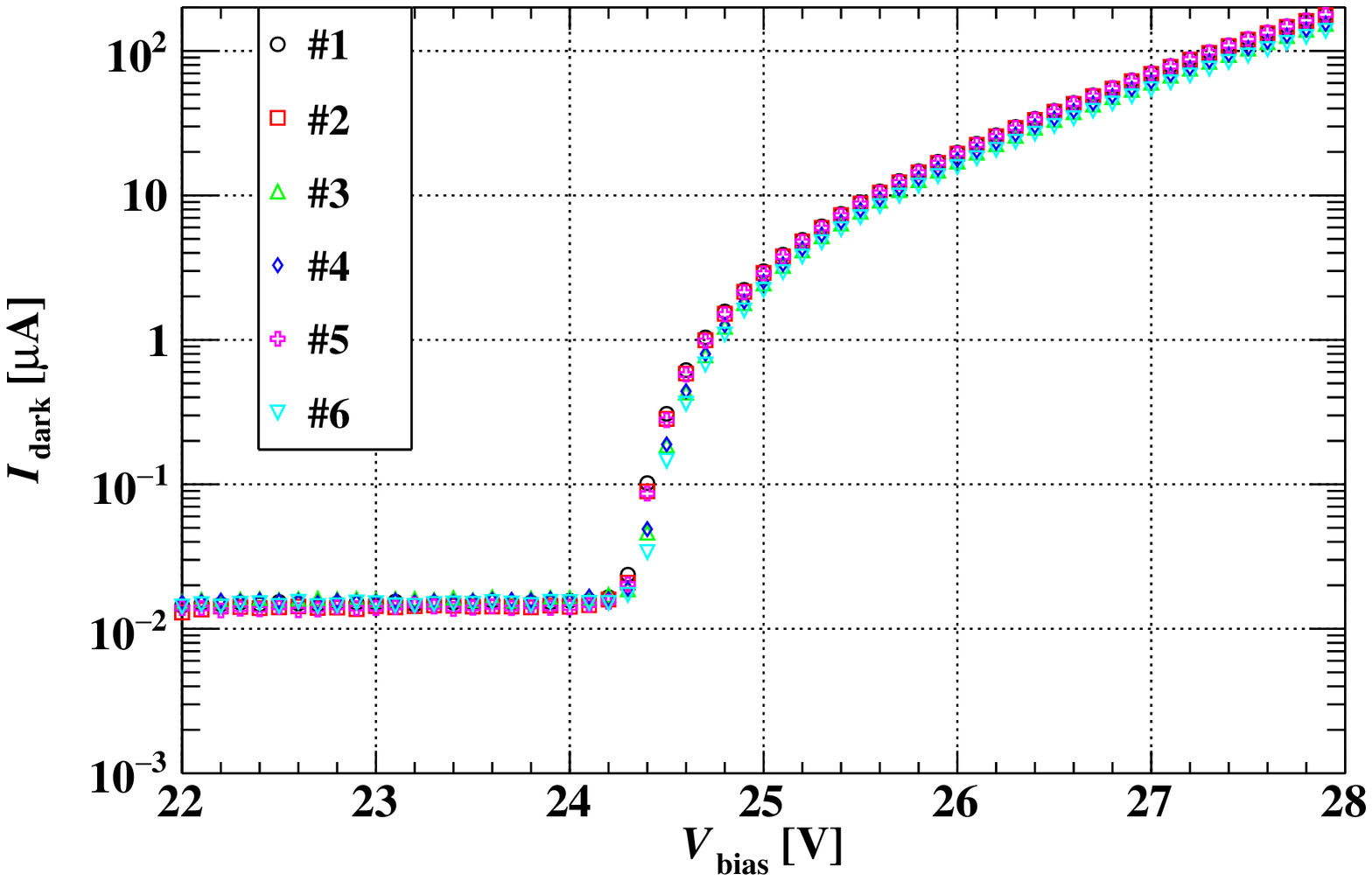}
		\subcaption{After 140 hours irradiation.}
	\end{minipage}
		\hfil
	\begin{minipage}[t]{\columnwidth}
		\centering
		\includegraphics[clip, width=0.9\columnwidth]{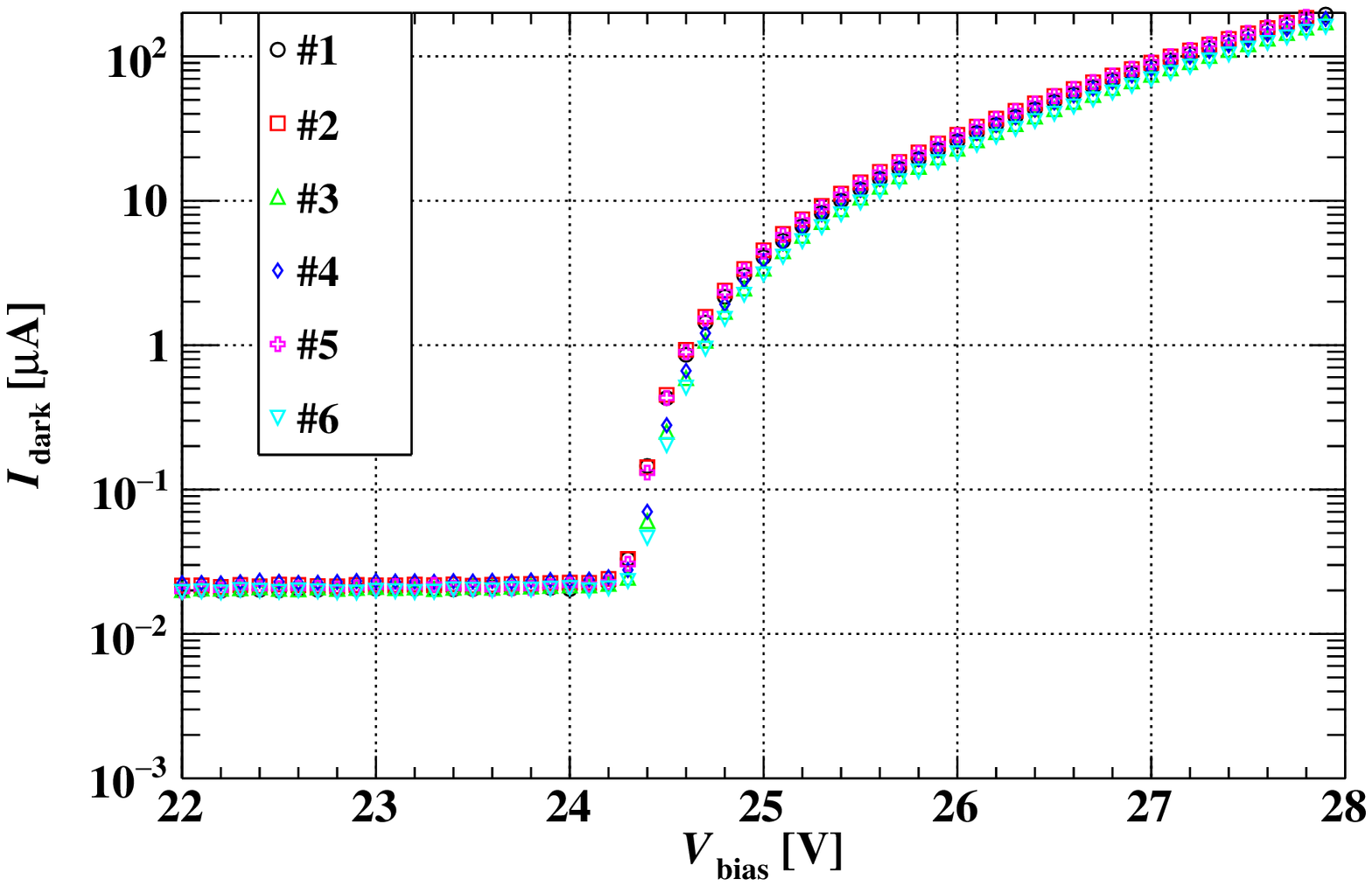}
		\subcaption{After 210 hours irradiation.}
	\end{minipage}
	
	\begin{minipage}[t]{\columnwidth}
		\centering
		\includegraphics[clip, width=0.9\columnwidth]{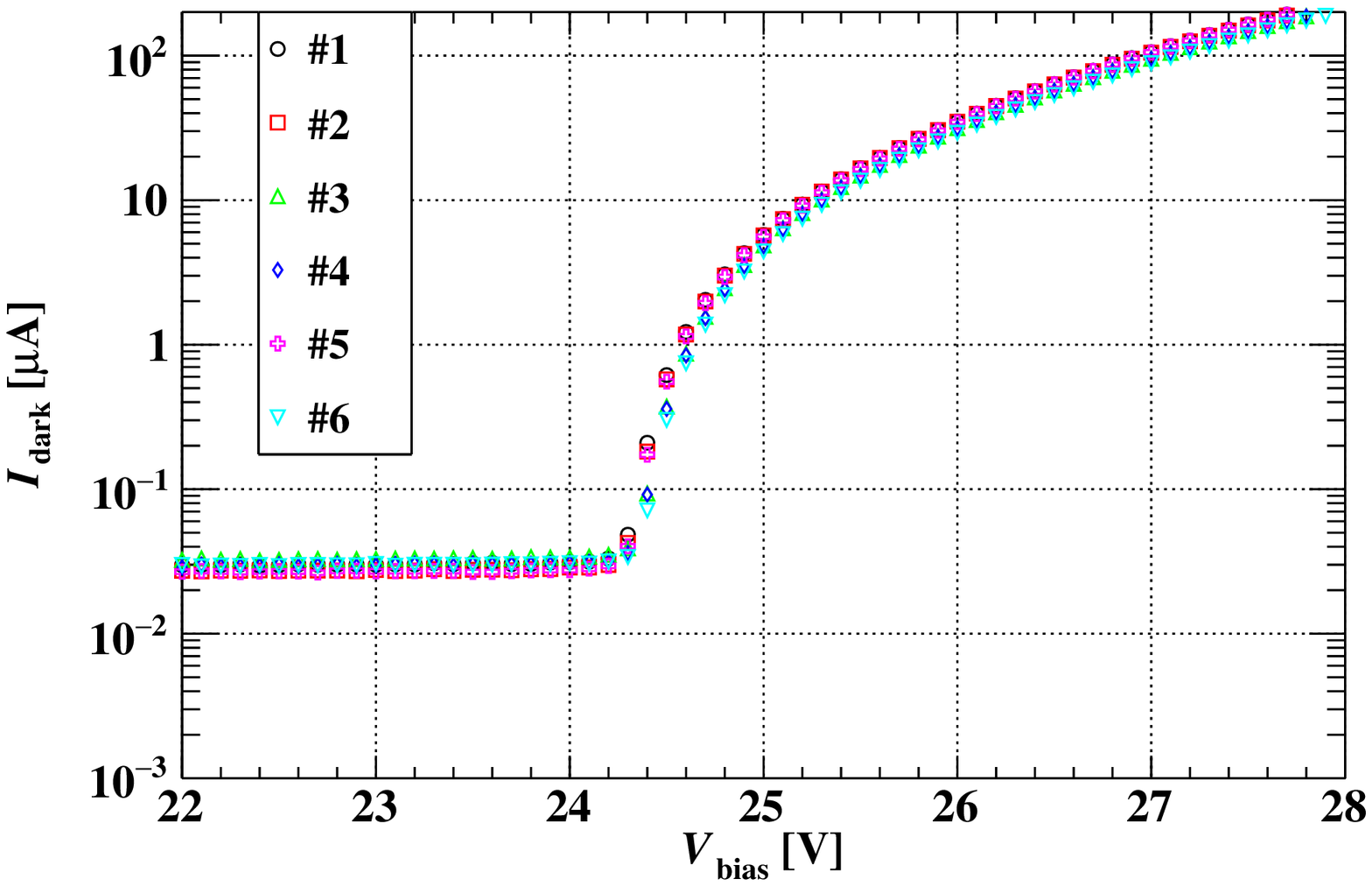}
		\subcaption{After 280 hours irradiation.}\label{Fig:280hour}
	\end{minipage}
	\hfil
	\begin{minipage}[t]{\columnwidth}
		\centering
		\includegraphics[width=0.9\columnwidth]{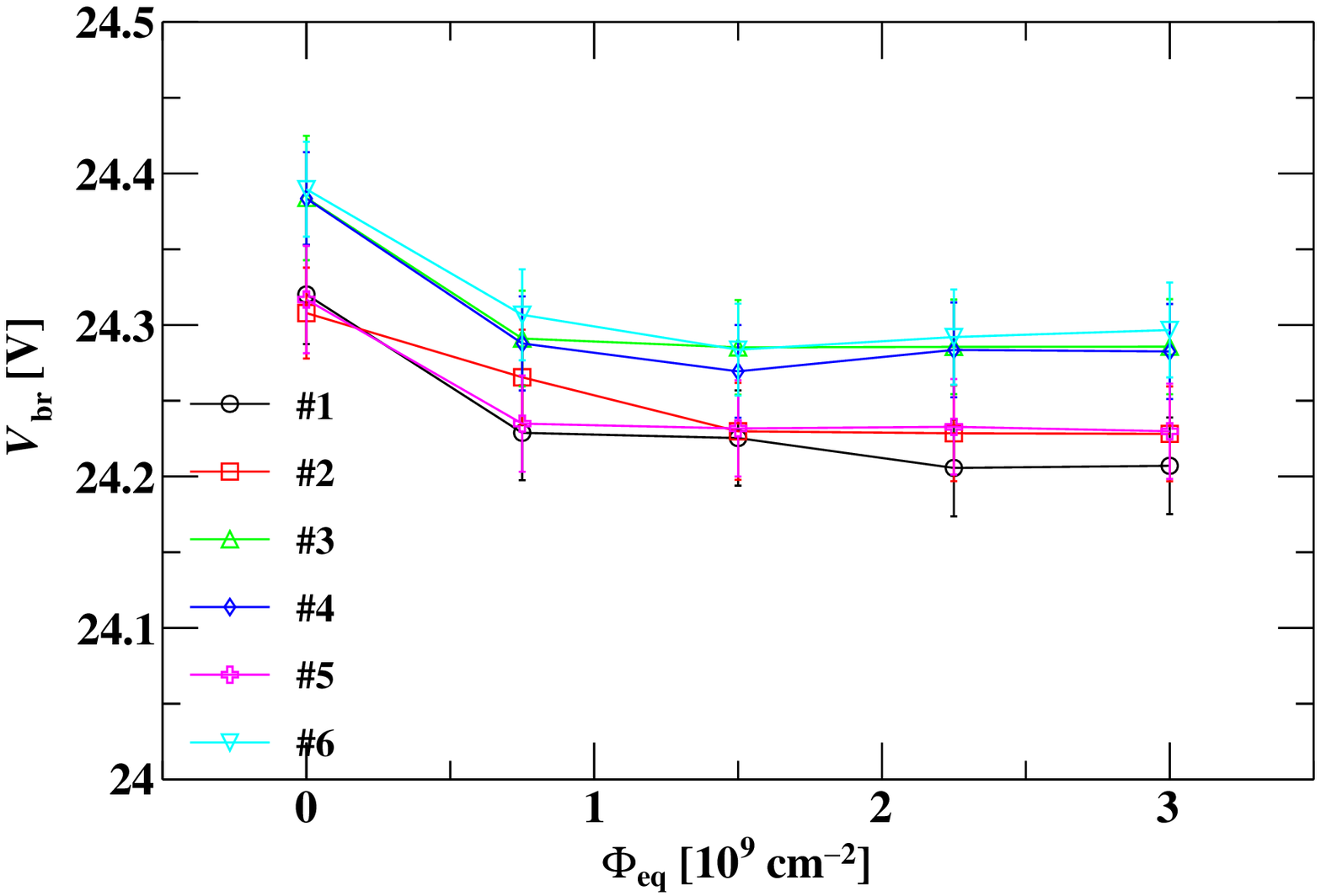}
		\subcaption{Breakdown voltages of single SiPMs versus fluence.}\label{Fig:breakdown_single}
	\end{minipage}

	\begin{minipage}[t]{\columnwidth}
		\centering
		\includegraphics[clip, width=0.9\columnwidth]{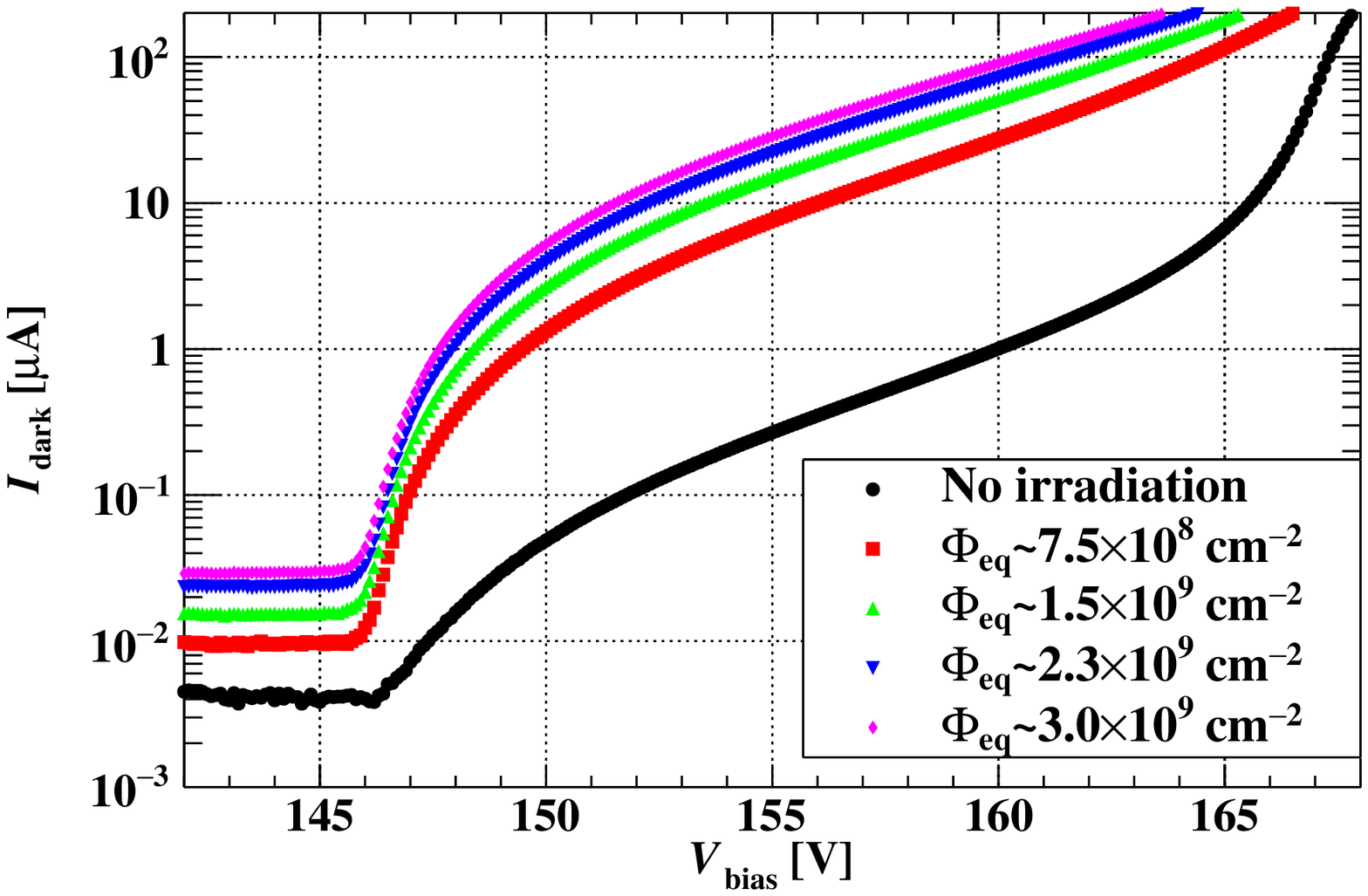}
		\subcaption{Six series-connected SiPMs at each irradiation step at 30$^\circ$C.}\label{Fig:IV_6series_30deg}
	\end{minipage}
	\hfil
	\begin{minipage}[t]{\columnwidth}
		\centering
		\includegraphics[clip, width=0.9\columnwidth]{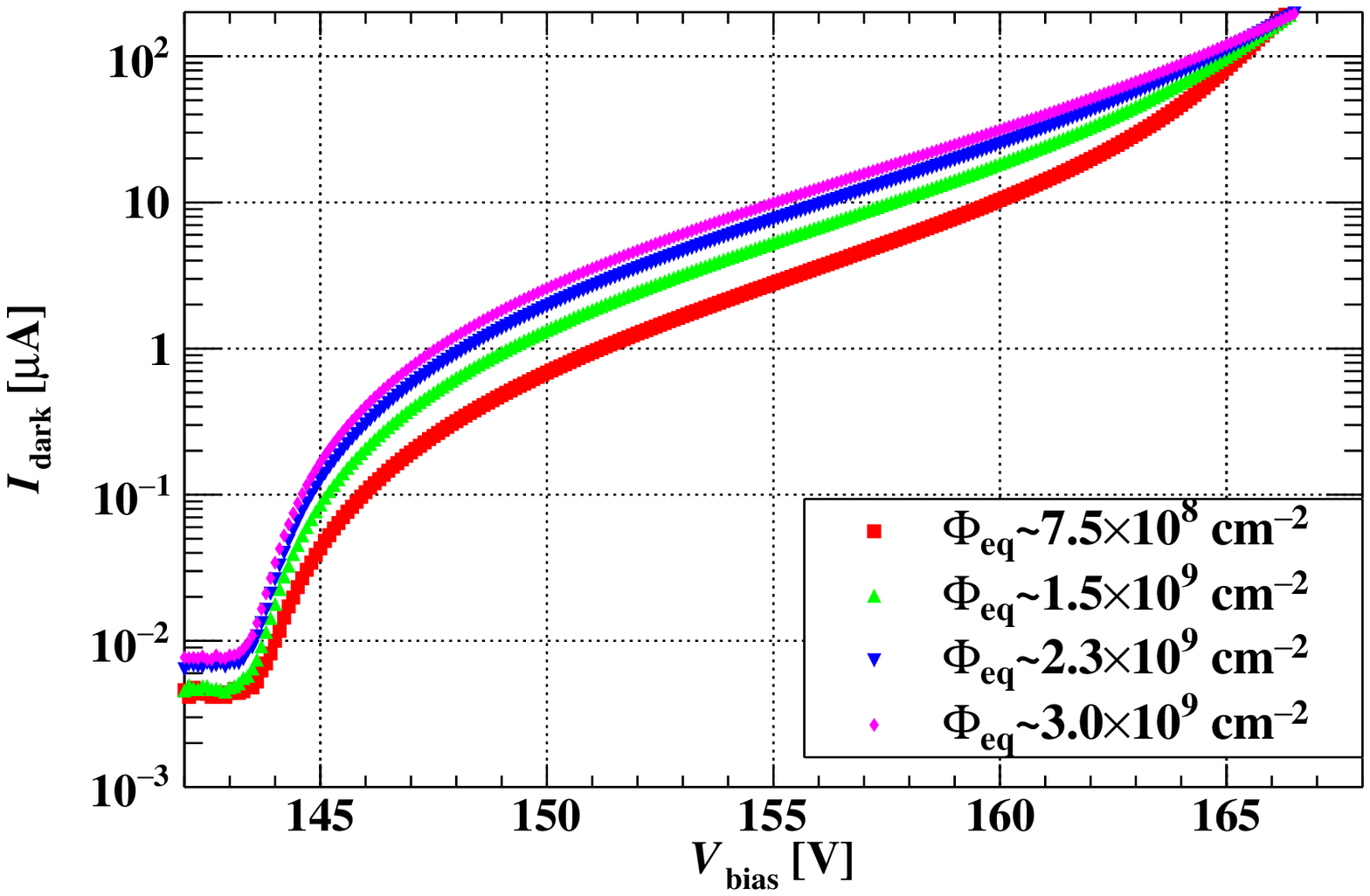}
		\subcaption{Six series-connected SiPMs at each irradiation step at 10$^\circ$C.}\label{Fig:IV_6series_10deg}
	\end{minipage}

	\caption{The I-V curves of the six electron-irradiated SiPMs. 
        The I-V data of each SiPM were taken at 30$^\circ$C (\subref{Fig:0hour}--\subref{Fig:280hour}), and the breakdown voltages are plotted in (\subref{Fig:breakdown_single}). 
        The I-V data of six series-connected SiPMs were taken at 30$^\circ$C 
        (\subref{Fig:IV_6series_30deg}) and 10$^\circ$C (\subref{Fig:IV_6series_10deg}).
        }\label{Fig:IV_curves}
\end{figure*}

\Frefs{Fig:0hour}{Fig:280hour} show the dark current $I_\mathrm{dark}$ versus reverse bias voltage $V_\mathrm{bias}$ (I-V) curves of each of the six electron-irradiated SiPMs and 
\frefs{Fig:IV_6series_30deg}{Fig:IV_6series_10deg} show the I-V 
curves when they are connected in series (taken at 0.1~V steps). 
The breakdown voltage $V_\mathrm{br}$ is calculated from the maximum point in the second-derivative of the logarithm of $I_\mathrm{dark}$ with three-point smoothing~\cite{simonetta} and the systematic uncertainties are estimated from the change by different smoothing. 
Table~\ref{Tab:Vbr_table} summarises the breakdown voltages and confirms that the sum of the single breakdown voltage values is consistent with the value from those connected in series when $I_{\mathrm{dark}}$ values are similar among the SiPMs. 
\begin{table}[tb]
  \centering
  \small
   \caption{$V_{\mathrm{br}}$ values before irradiation at 30$^\circ$C.}
   \label{Tab:Vbr_table}
    \begin{tabular}{cc} \hline
    SiPM & $V_{\mathrm{br}}$ (V)\\ \hline \hline
      \#1 & 24.32 $\pm$ 0.03 \\
      \#2 & 24.31 $\pm$ 0.03 \\
      \#3 & 24.38 $\pm$ 0.04 \\
      \#4 & 24.38 $\pm$ 0.03 \\
      \#5 & 24.32 $\pm$ 0.04 \\
      \#6 & 24.39 $\pm$ 0.03 \\
      Sum of \#1 - \#6 $V_{\mathrm{br}}$ & 146.1 $\pm$ 0.2 \\
      \#1 - \#6 Connected in Series & 146.21 $\pm$ 0.05 \\
      \hline
    \end{tabular}
\end{table}

The dark currents $I_\mathrm{dark}$ increase significantly under electron irradiation; roughly by one order of magnitude below the breakdown voltage and two orders of magnitude above it.
On the other hand, the breakdown voltages are only marginally affected by the electron-irradiation damage as shown in
\fref{Fig:breakdown_single}. 
The systematic difference between non-irradiated and irradiated points could be due to the algorithm 
for the $V_\mathrm{br}$ determination because the slope of the current after the breakdown becomes steeper after the first irradiation.

\subsection{Waveforms}

\begin{figure}[tb]
		\centering
		\includegraphics[clip, width=0.99\columnwidth]{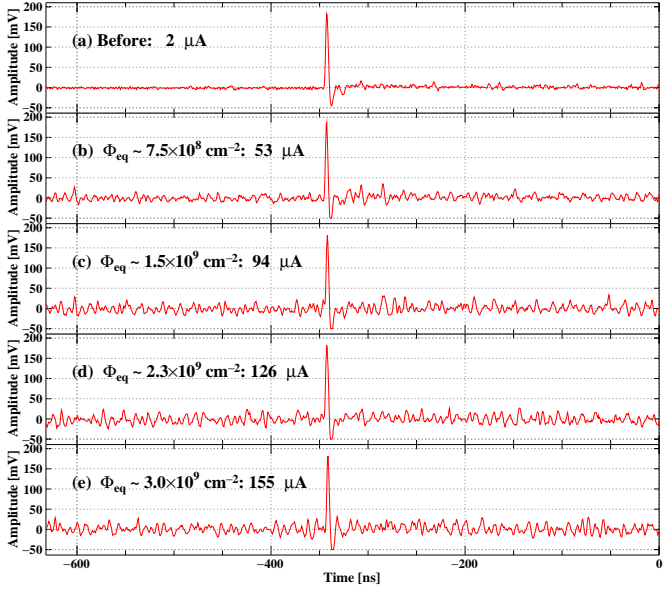}
		\caption{Examples of waveforms
               at different irradiation steps at 30$^\circ$C. The applied voltage was set to a fixed value 
              (162.5 V). The dark current values obtained from the I-V curves are also shown.}
              \label{Fig:Waveform_Example}
\end{figure}

\begin{figure}[tb]
		\centering
		\includegraphics[clip, width=0.99\columnwidth]{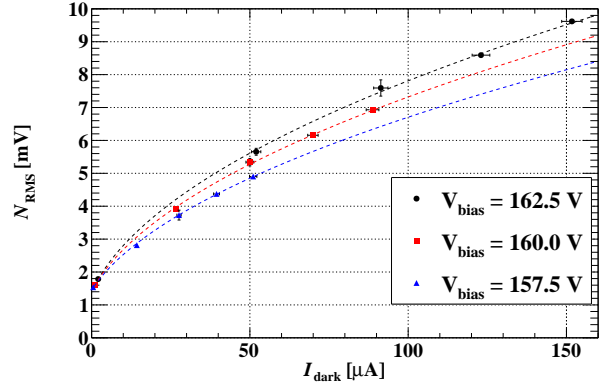}
		\caption{Relation between the dark current and the RMS noise at $30^{\circ}$C for three applied voltages.
		The dashed curves $N_\mathrm{RMS} = \sqrt{\alpha I_\mathrm{dark} + N_0^2}$ are fitted to the data.
              }\label{Fig:CurrentVSNoise}
\end{figure}

\Fref{Fig:Waveform_Example} shows an example of the waveform from the series-connected SiPMs with the electron-irradiated samples at each irradiation step. 
The fluctuation of the baseline increased significantly as shown in \fref{Fig:CurrentVSNoise}, where $N_\mathrm{RMS}$ is the mode value of RMS noise calculated with 
baseline points before the signal region. 

\subsection{Timing resolution}
\label{sec:timing_resolution}

\begin{figure}[tb]
		\centering
		\includegraphics[clip, width=0.85\columnwidth]{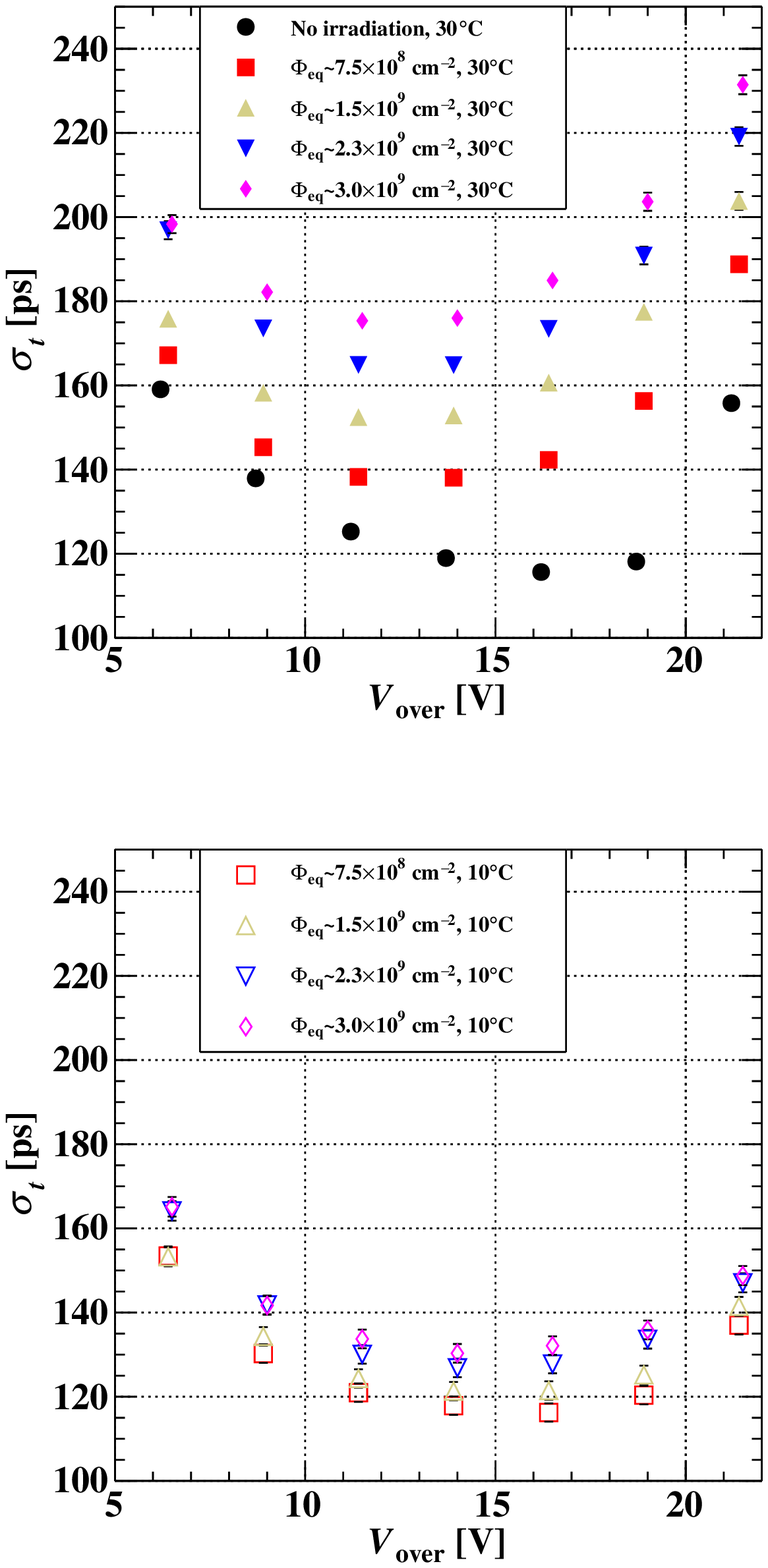}
		\caption{Timing resolution versus over-voltage at 30$^\circ$C (top) and 10$^\circ$C (bottom). 
              The CFD fraction is 20\%.}\label{Fig:BiasScan}
\end{figure}

\begin{figure}[tb]
		\centering
		\includegraphics[clip, width=0.98\columnwidth]{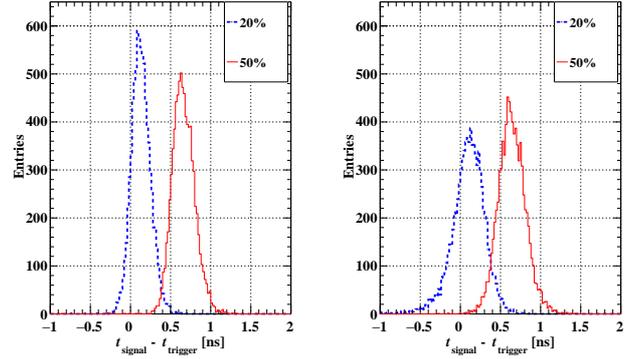}
		\caption{Distribution of the obtained timing $t_{\mathrm{signal}} - t_{\mathrm{trigger}}$, (left) before irradiation and (right) after 280-hour irradiation with different CFD fractions at $V_\mathrm{over} \approx 16.5$~V (30$^\circ$C).
              }\label{Fig:TimeDistribution}
\end{figure}

\begin{figure}[tb]
		\centering
		\includegraphics[clip, width=0.9\columnwidth]{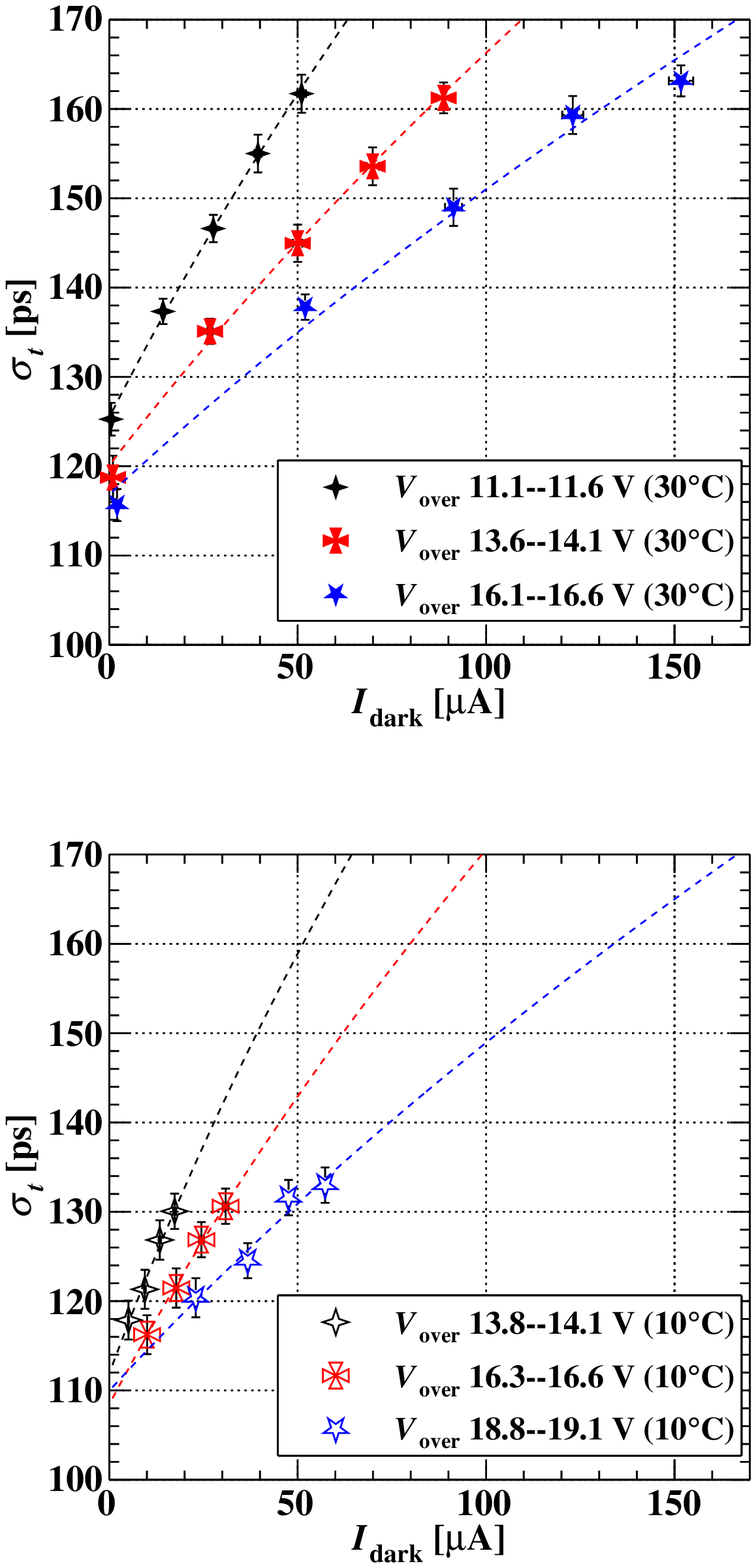}
		\caption{Relation between the dark current and the timing resolution at 30$^\circ$C (top) and 10$^\circ$C (bottom). 
              The CFD fraction was optimised in the range of 10--60\%. 
              The dashed curves $\sigma_t=\sqrt{\beta I_\mathrm{dark} + \sigma_\mathrm{0}^2}$ are fitted to the data.
              }\label{Fig:CurrentScale}

\end{figure}

\begin{figure}[tb]
		\centering
		\includegraphics[clip, width=0.9\columnwidth]{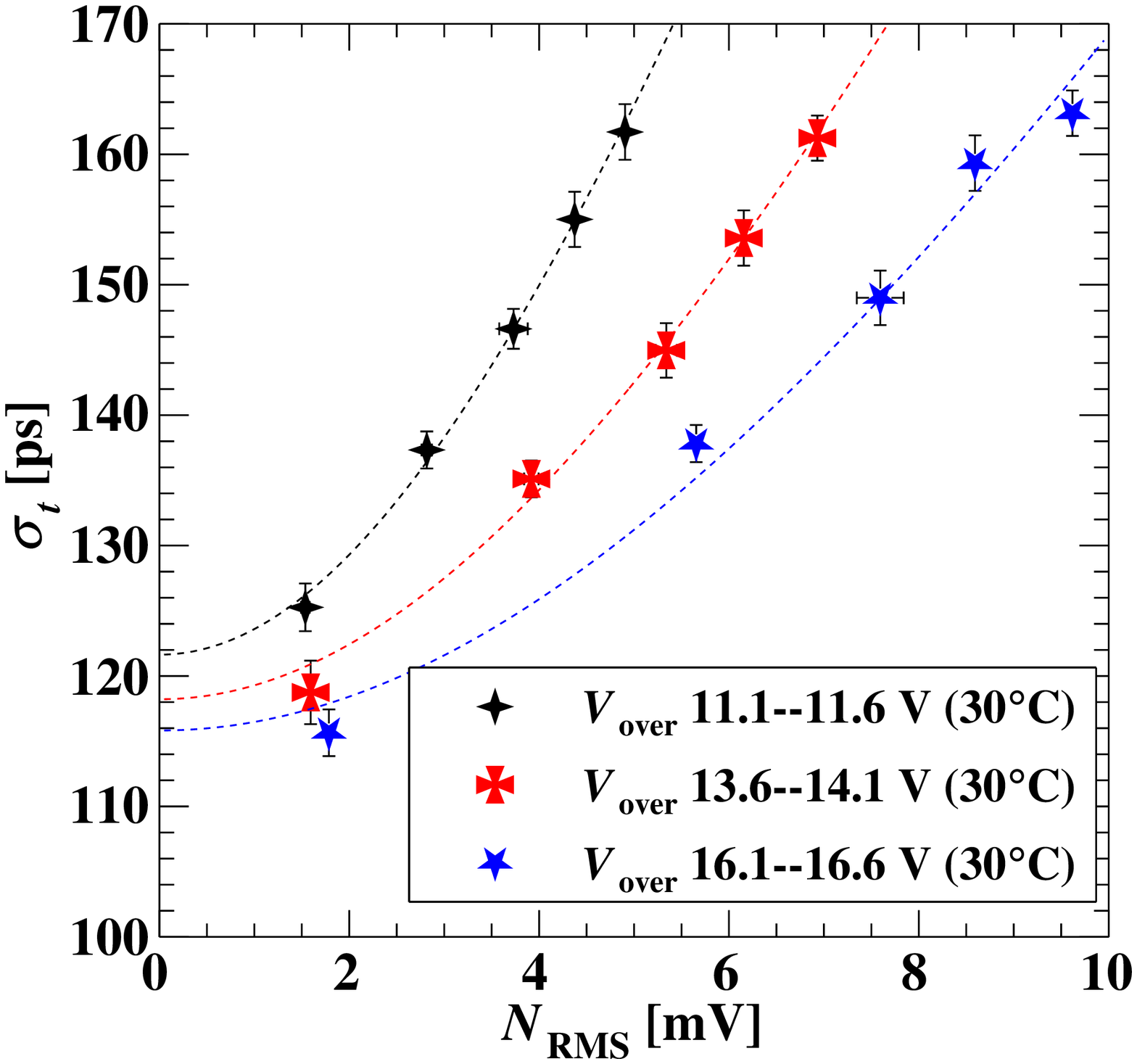}
		\caption{Relation between the RMS noise and the timing resolution at $30^{\circ}$C for three applied voltages.
		The dashed curves are $\sigma_t = \sqrt{\gamma N_\mathrm{RMS}^2 + \sigma_\mathrm{0}^2}$ fitted to the data.
              }\label{Fig:ResolutionVSNoise}
\end{figure}

The timing resolution $\sigma_t$ is optimised scanning the bias voltage in the range $V_\mathrm{bias} = 
150 - 166$ V ($V_\mathrm{over} \equiv V_\mathrm{bias} - V_\mathrm{br} = 6 - 22$~V), in steps of 2.5 V both at 30$^\circ$C and 10$^\circ$C.
The data were analysed with a fixed CFD fraction of 20\% and the results are shown in \fref{Fig:BiasScan}. 
The resolution first improves with $V_\mathrm{bias}$ due to a higher gain and PDE, 
but at some voltage starts to deteriorate due to increased dark counts.
As the fluence increases, 
the optimal voltage, where $\sigma_t$ reaches the minimum, shifts to lower voltages. 
$\sigma_t$ at each optimal voltage gradually worsens and after the fourth irradiation the degradation reaches $\sim50$\%.  

Next, three points around each optimal voltage ($V_{\mathrm{bias}}$ = 157.5, 160.0, 162.5~V) are selected, $V_\mathrm{over} \approx 11.5$, 14.0, 16.5~V 
for the 30$^\circ$C datasets and $V_\mathrm{over} \approx 14.0$, 16.5, 19.0~V for the 10$^\circ$C datasets, 
and the CFD fraction is optimised in steps of 10\% from 10\% to 60\% for each dataset. \Fref{Fig:TimeDistribution} shows the distribution of the calculated timing $t_{\mathrm{signal}} - t_{\mathrm{trigger}}$ after irradiation with different CFD fractions at $V_\mathrm{over} \sim 16.5$~V (30$^\circ$C).
As the noise increases, the optimal fraction value becomes higher: from 20\% to 50\%. 

\Fref{Fig:CurrentScale} shows the relation between $\sigma_t$ and $I_\mathrm{dark}$.
When the dark current increases to $I_\mathrm{dark}\sim 100$ $\mathrm{\muup A}$, 
$\sigma_t$ deteriorates by $\sim 30$\% at $V_\mathrm{over}\sim 16.5$~V. 

Compared to the 30$^\circ$C case, $I_\mathrm{dark}$ is significantly lower at 10$^\circ$C 
at each irradiation step. 
Comparing the two cases at $V_\mathrm{over}\sim 16.5$~V after 
140-hour irradiation (the third black marker at 30$^\circ$C and the second white marker at 10$^\circ$C in the plot),
the dark current decreases from $I_\mathrm{dark}\sim 90~\mathrm{\muup A}$ to $\sim20~\mathrm{\muup A}$ and the
deterioration of the timing resolution reduces from $29$\% to $5$\%.

The $\sqrt{I_{\mathrm{dark}}}$ dependence of $\sigma_t$ can be explained by the relation between $N_{\mathrm{RMS}}$ and $I_{\mathrm{dark}}$. 
The signal height was constant during these measurements within the range of the measurement errors ($A \sim$ 0.15 -- 0.16 V from the most-probable value of its Landau-like distribution).
\Fref{Fig:ResolutionVSNoise} shows the relation between $\sigma_t$ and $N_{\mathrm{RMS}}$.
$\sigma_{t}$ can be expressed by the square root of the square sum of the intrinsic timing resolution ($\sigma_{0}$) and the additional fluctuation from the noise ($N_{\mathrm{RMS}}$).
Also, $N_{\mathrm{RMS}}$ can be scaled by the $\sqrt{I_{\mathrm{dark}}}$ dependence
as shown in \fref{Fig:CurrentVSNoise}. 
Hence the deterioration of $\sigma_t$ can be described by $\sqrt{I_{\mathrm{dark}}}$ dependence.

\section{Series connection of differently damaged SiPMs}
\label{sec:differently-damaged-measurement}
The previous section presents the relation between $I_{\mathrm{dark}}$ and 
$\sigma_t$ of a scintillator counter with equally-irradiated series-connected SiPMs. 
Under more realistic assumptions, SiPMs may be irradiated with different fluences due to 
position dependence of the particle flux.
In the following, results with differently-irradiated SiPMs connected in series are presented.

\subsection{Patterns of SiPM combination}

\begin{figure}[btp]
	\centering
	\includegraphics[width=0.5\columnwidth]{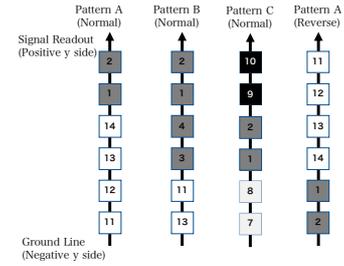}
	\caption{Schematic of SiPMs' pattern A, pattern B, pattern C in normal order and pattern A in reverse order. The SiPMs with highest dark currents are in black, those with lowest in white.}
	\label{Fig:AlignmentIllust}
\end{figure}

The combinations of SiPMs used for this 
test are:\footnote{
Here the SiPMs are distinguished not by the kind of irradiating 
particles but only by the current level.}
\begin{itemize}
\item Pattern A is composed of two electron-irradiated ($\Phi_\mathrm{eq} \approx 3\times 10^{9}~\mathrm{cm^{-2}}$) SiPMs (\#1 and \#2 [or \#5]\footnote{SiPM \#2 
was broken during the measurement and was replaced with \#5, 
which has the same damage level and I-V characteristics as \#2.}) 
 and four non-irradiated SiPMs (\#11--\#14).
\item Pattern B is composed of four electron-irradiated ($\Phi_\mathrm{eq} \approx 3\times 10^{9}~\mathrm{cm^{-2}}$) SiPMs (\#1--\#4 [\#5]) and two non-irradiated SiPMs (\#11, \#13).
\item Pattern C is composed of two electron-irradiated SiPMs (\#1, \#2 [\#5]) and four neutron-irradiated SiPMs, two with the fluence level of  $\Phi_\mathrm{eq} \approx 8.7 \times 10^8~\mathrm{cm^{-2}}$ samples (\#7, \#8) and the other two with $\Phi_\mathrm{eq} \approx 5.5 \times 10^9~\mathrm{cm^{-2}}$ ones (\#9, \#10).
\item Pattern D is composed of six electron-irradiated ($\Phi_\mathrm{eq} \approx 3\times 10^{9}~\mathrm{cm^{-2}}$) SiPMs (\#1--\#6). They are the same SiPMs used in Sect.~\ref{sec:equally-damaged-measurement}.

\end{itemize}
Pattern A and pattern B are considered as extreme cases to highlight the effects of differential 
irradiation, while pattern C simulates a gradient damage level closer to a realistic condition.

The coordinate system on the tile is defined such that 
the $x$-axis runs along the long direction while
the $y$-axis runs along the short direction 
with the origin point at the centre of the tile. 
The SiPMs under test are located at the negative $x$ side; the signal is extracted from the positive $y$ side, while the negative $y$ side is connected to the ground line.
The more highly-damaged SiPMs were located at larger $y$ position: 
this configuration of SiPMs is called ``normal order".
Data were taken also with SiPMs ordered in the opposite way along the $y$ direction: 
this configuration is called ``reverse order". \Fref{Fig:AlignmentIllust} illustrates the layout of each pattern.

\subsection{I-V curves and apparent breakdown voltage shift}

\begin{figure*}[p]
      	\begin{minipage}[t]{0.45\linewidth}
      		\centering
		\includegraphics[clip, width=1.\columnwidth]{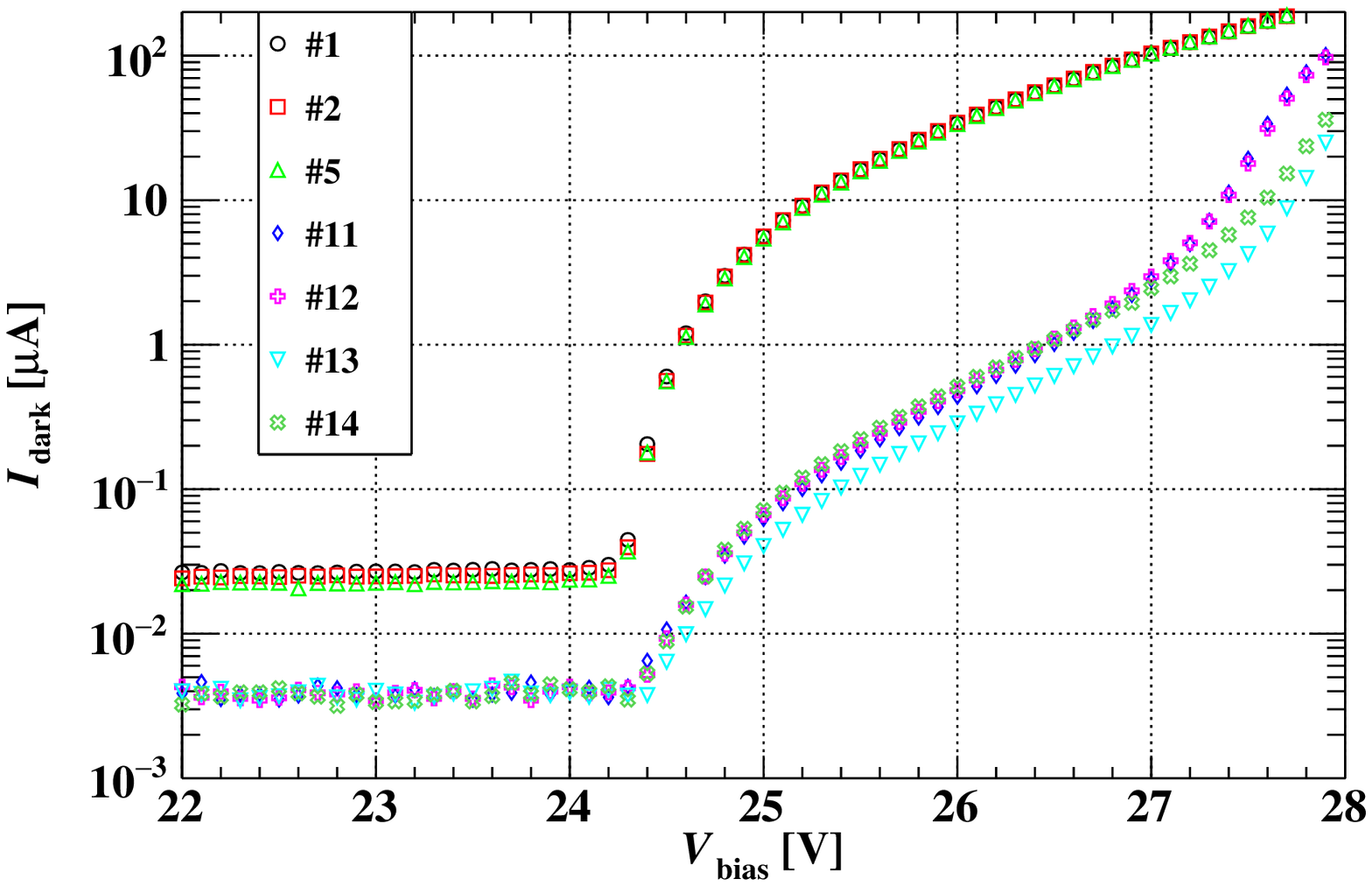}
		\caption{I-V curves of each SiPM used for pattern A.}
		\label{IV_Pattern_A}
	\end{minipage} 
      	\begin{minipage}[t]{0.45\linewidth}
		\centering
		\includegraphics[clip, width=1.\columnwidth]{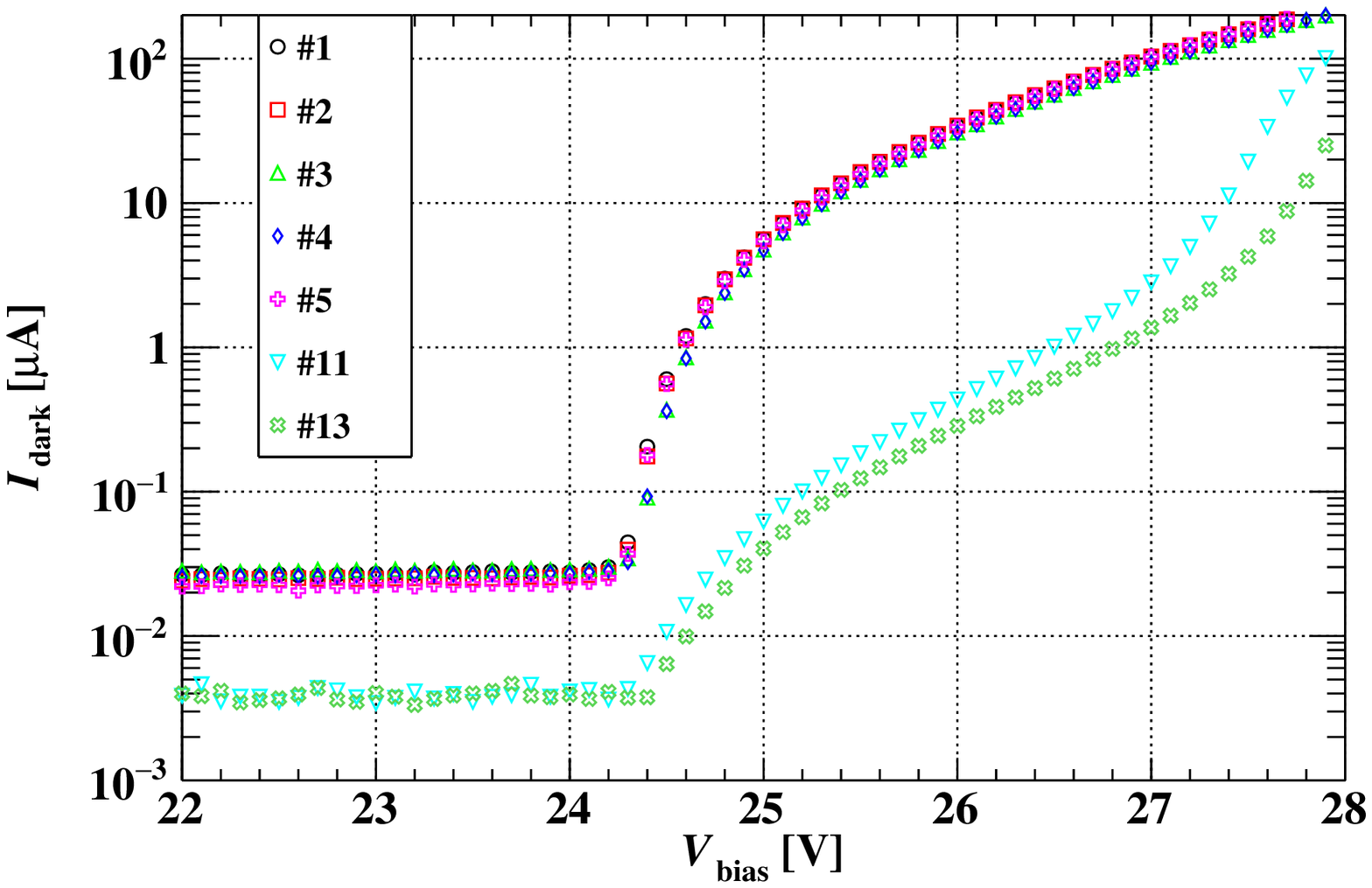}
		\caption{I-V curves of each SiPM used for pattern B}
		\label{IV_Pattern_B}
	\end{minipage}
	\begin{minipage}[t]{0.45\linewidth}	
		\centering
		\includegraphics[clip, width=1.\columnwidth]{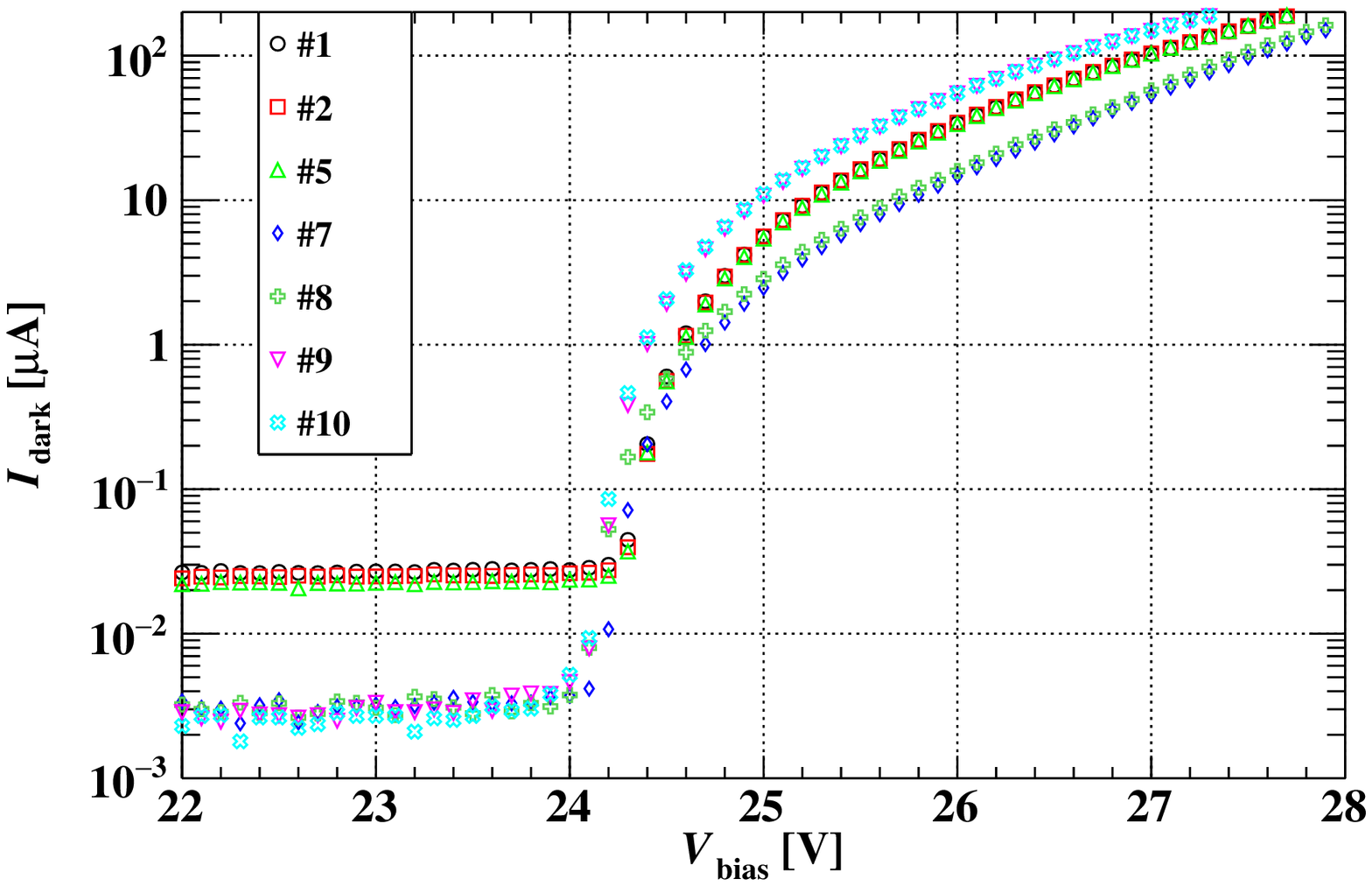}
		\caption{I-V curves of each SiPM used for pattern C.}
		\label{IV_Pattern_C}
	\end{minipage}		
	\begin{minipage}[b]{0.45\linewidth}
		\centering
		\includegraphics[clip, width=1.\columnwidth]{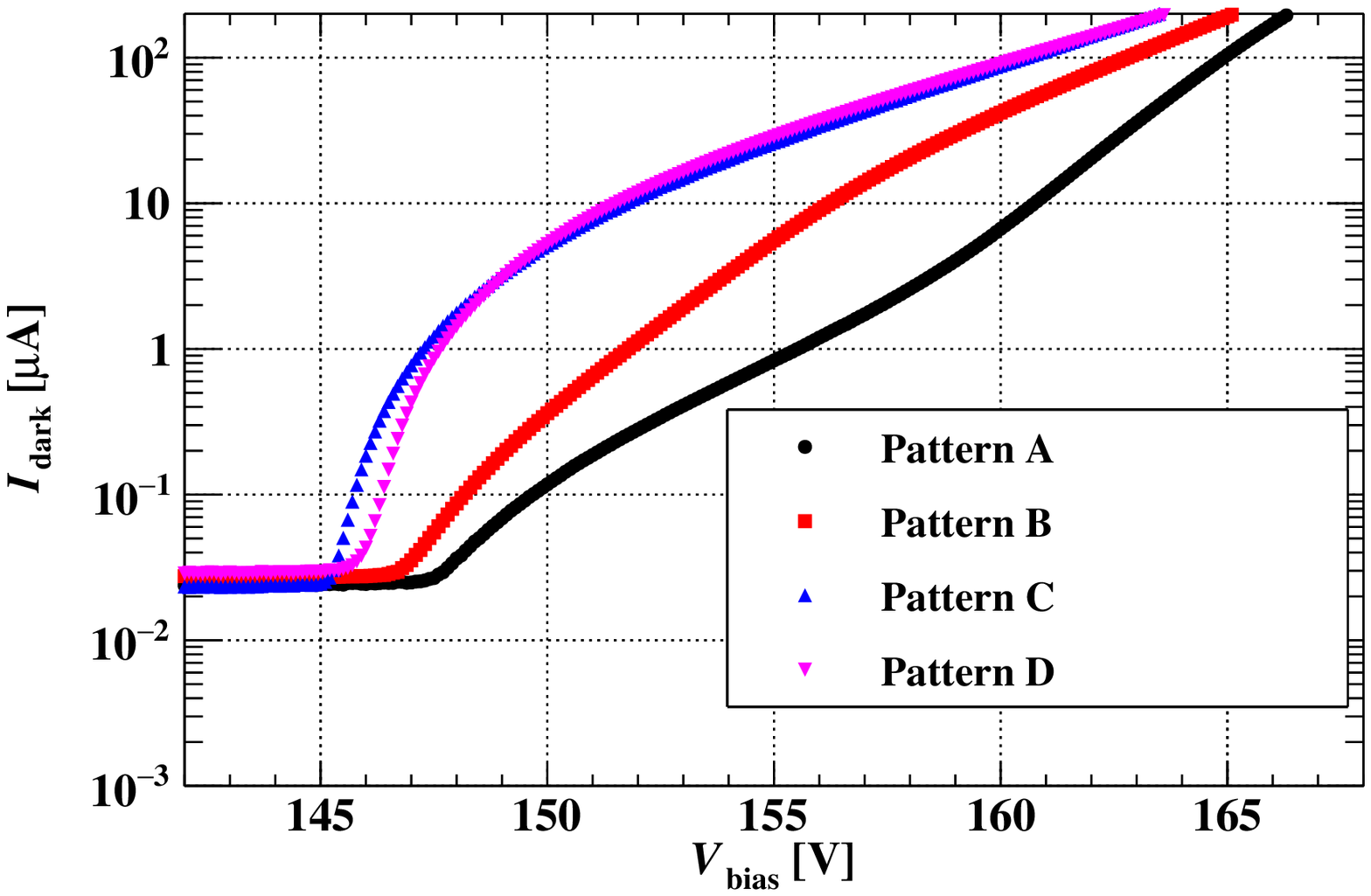}
		\caption{I-V curves of several patterns of series-connected SiPMs measured at 30$^\circ$C.}
		\label{IV_Unbalance_30deg}
	\end{minipage}
	\begin{minipage}[b]{0.45\linewidth}
		\centering
		\includegraphics[clip, width=1.\columnwidth]{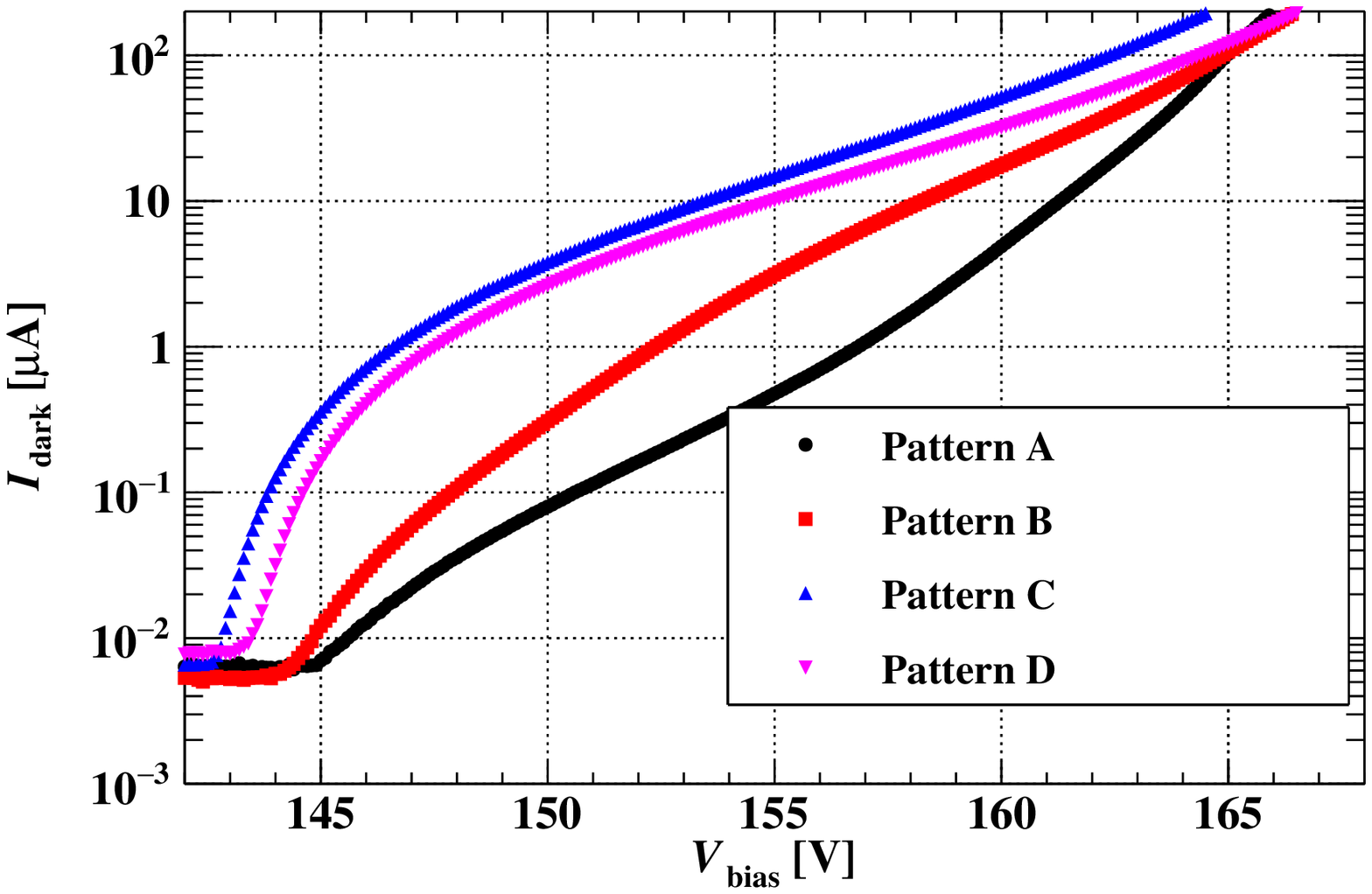}
		\caption{I-V curves of several patterns of series-connected SiPMs measured at 10$^\circ$C.}
		\label{IV_Unbalance_10deg}
	\end{minipage}
	\begin{minipage}{0.1\linewidth}
		\hspace{1mm}
	\end{minipage}
	\begin{minipage}[t]{0.45\linewidth}
		\centering
		\includegraphics[clip, width=1.\columnwidth, trim = 750 375 750 375]{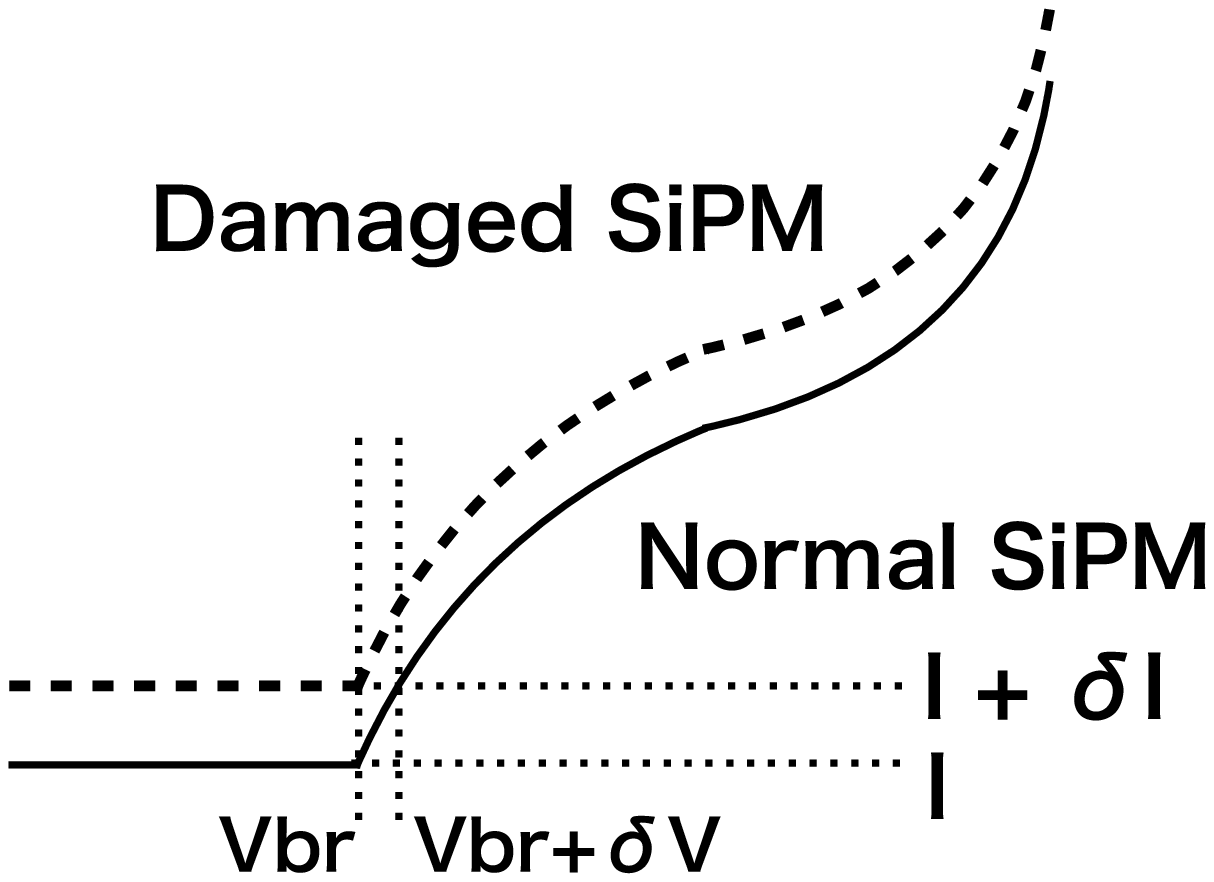}
		\caption{The explanation of breakdown voltage shift arising from the difference in dark currents. 
              If $n$ SiPMs have the same dark current $I$ at the breakdown voltage $V_\mathrm{br}$,
              the breakdown voltage of the series-connected SiPMs is expected to be $n\times V_\mathrm{br}$. 
              In pattern A, the breakdown voltage becomes 
              $6\times V_\mathrm{br} + 4 \times \delta V_\mathrm{br}$, and in pattern B, 
              $6\times V_\mathrm{br} + 2 \times \delta V_\mathrm{br}$ to accommodate the same current $I + \delta I$ 
              at the breakdown voltage.}\label{BreakdownShift_Model}
	\end{minipage}
\end{figure*}

\Frefss{IV_Pattern_A}{IV_Pattern_C} show the I-V curves of individual SiPMs used in patterns A, B, and C.
\Frefsa{IV_Unbalance_30deg}{IV_Unbalance_10deg} show the I-V 
curves  of them in series measured at 30$^\circ$C and 10$^\circ$C, respectively.
The breakdown voltages of patterns A and B turned out to be shifted to values higher than that of pattern D.
Table~\ref{Tab:Vbr_shift_table} shows the breakdown voltage values of pattern A , pattern D, and the SiPMs used for those patterns.

When the I-V characteristics of series-connected SiPMs differ, the applied voltage to each SiPM 
is adjusted to accommodate a common current.
As a consequence, the over-voltages for the six SiPMs differ from one another; even when the total applied voltage is below the apparent breakdown voltage,
 voltages applied to
non-damaged SiPMs may result to be higher than their breakdown voltages (positive over-voltages) to accommodate the same dark current 
as that of the damaged SiPMs below the breakdown (negative over-voltages). 
This mechanism causes an apparent breakdown voltage shift depicted in \Fref{BreakdownShift_Model}.

%
\begin{table}[tb]
  \centering
  \small
   \caption{$V_{\mathrm{br}}$ values after the full-period electron irradiation.}
   \label{Tab:Vbr_shift_table}
    \begin{tabular}{cc} \hline
    SiPM & $V_{\mathrm{br}}$ (V)\\ \hline \hline
      \#1 (30$^\circ$C) & 24.20 $\pm$ 0.03 \\
      \#2 (30$^\circ$C) & 24.23 $\pm$ 0.03 \\
      \#3 (30$^\circ$C) & 24.29 $\pm$ 0.03 \\
      \#4 (30$^\circ$C) & 24.28 $\pm$ 0.03 \\
      \#5 (30$^\circ$C) & 24.23 $\pm$ 0.03 \\
      \#6 (30$^\circ$C) & 24.30 $\pm$ 0.03 \\
      \#11 (30$^\circ$C) & 24.28 $\pm$ 0.03 \\
      \#12 (30$^\circ$C) & 24.33 $\pm$ 0.03 \\
      \#13 (30$^\circ$C) & 24.39 $\pm$ 0.03 \\
      \#14 (30$^\circ$C) & 24.32 $\pm$ 0.03 \\
      Sum of \#1 - \#6  & 145.5 $\pm$ 0.2 \\
      Sum of \#1, \#2, \#11 - \#14 & 145.7 $\pm$ 0.2 \\
      Pattern D (30$^\circ$C) & 145.94 $\pm$ 0.05 \\
      Pattern A (30$^\circ$C) & 147.51 $\pm$ 0.05 \\
      Pattern D (10$^\circ$C) & 143.43  $\pm$ 0.05 \\
      Pattern A (10$^\circ$C) & 144.89 $\pm$ 0.05 \\
      \hline
    \end{tabular}
\end{table}

\subsection{Bias voltage scan at centre}
\label{sec:scan_center}
\begin{figure}[t]
		\centering
		\includegraphics[clip, width=1.\columnwidth]{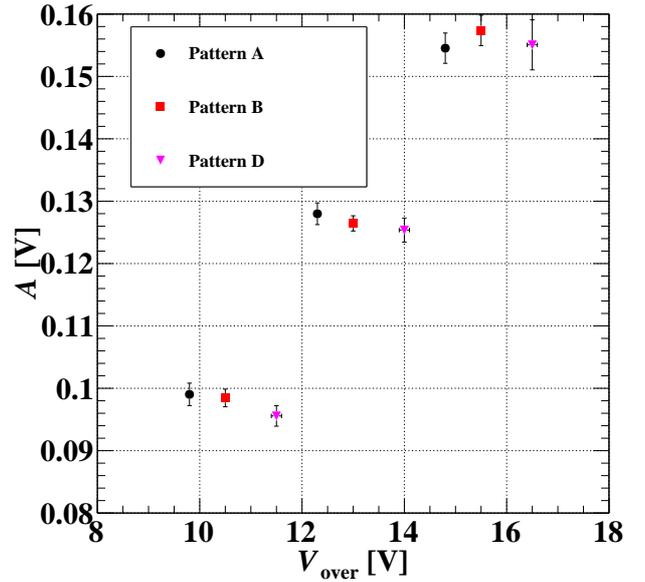}
		\caption{Pulse height at three common voltages (157.5 V, 160.0 V, and 162.5 V)  for different SiPM-combinations
              taken at 30$^\circ$C. Though the over-voltages from the apparent breakdown voltages are smaller for patterns A and B,
              the pulse heights are the same in the three configurations. 
              Pattern C is not shown here because the neutron-irradiated SiPMs have different breakdown voltages from the others.}
             \label{Bias_Scan_Unbalance_Height}
\end{figure}
\begin{figure}[t]
      		\centering
		\includegraphics[clip, width=1.\columnwidth]{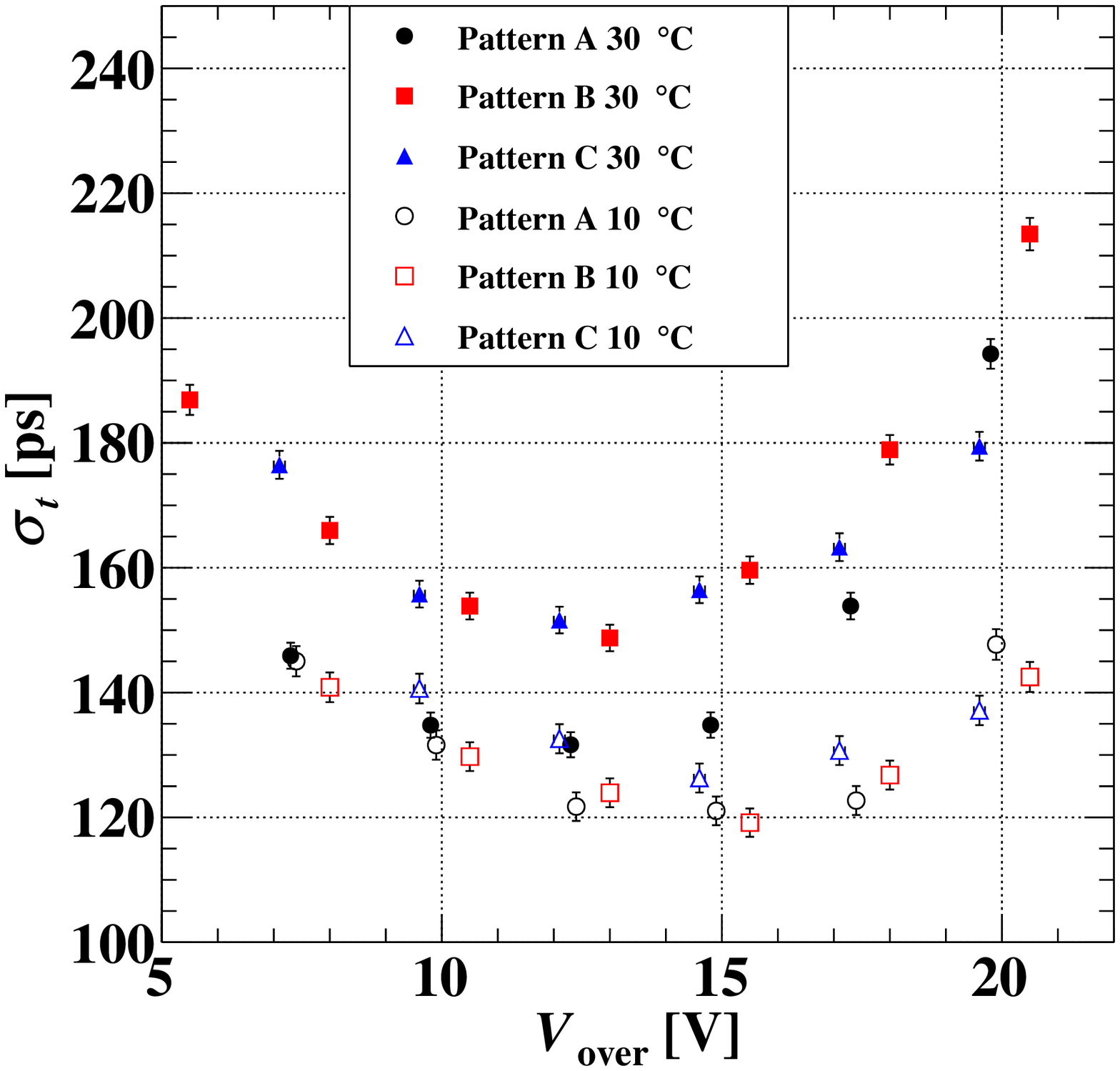}
		\caption{Timing resolution as a function of the over-voltage for the three SiPM-patterns taken at 
              30$^\circ$C and 10$^\circ$C. The CFD fraction was fixed to 20\%}\label{Bias_Scan_Unbalance_Resolution}
\end{figure}

First, from the over-voltage scan data with the measurement setup presented in Sect.~\ref{sec:measurements}
the pulse height $A$ and the timing resolution $\sigma_t$ were evaluated. 
\Frefsa{Bias_Scan_Unbalance_Height}{Bias_Scan_Unbalance_Resolution} shows the results.
The pulse height is defined as the most-probable value of its Landau-like distribution.
Although over-voltages from the apparent breakdown voltages become smaller, the pulse heights remain constant as in pattern D; 
that is, the total effective over-voltages remain constant.
The timing resolution is recovered again by cooling.

\subsection{Hit position dependence in a counter}

Next, the dependence of various variables on hit position in counters was examined.\footnote{In this subsection, the pulse height and the noise situation were changed from the measurements in Sect.~\ref{sec:timing_resolution} and \ref{sec:scan_center} due to a change in the setup. However, those parameters were stable during each measurement, and so the position dependence in different damage patterns can directly be compared.} 
Data were taken at different positions in the range of $-4.25 \leq x \leq 4.25~\mathrm{cm}$ and $-1 \leq y \leq 1~\mathrm{cm}$ in the normal order.
Then, the measurements were repeated in the reverse order to clearly distinguish the effect of radiation damage from other effects.
\subsubsection*{Pulse height}

\begin{figure*}[t]
	\begin{minipage}[t]{\columnwidth}
	\centering
	\includegraphics[width=\columnwidth]{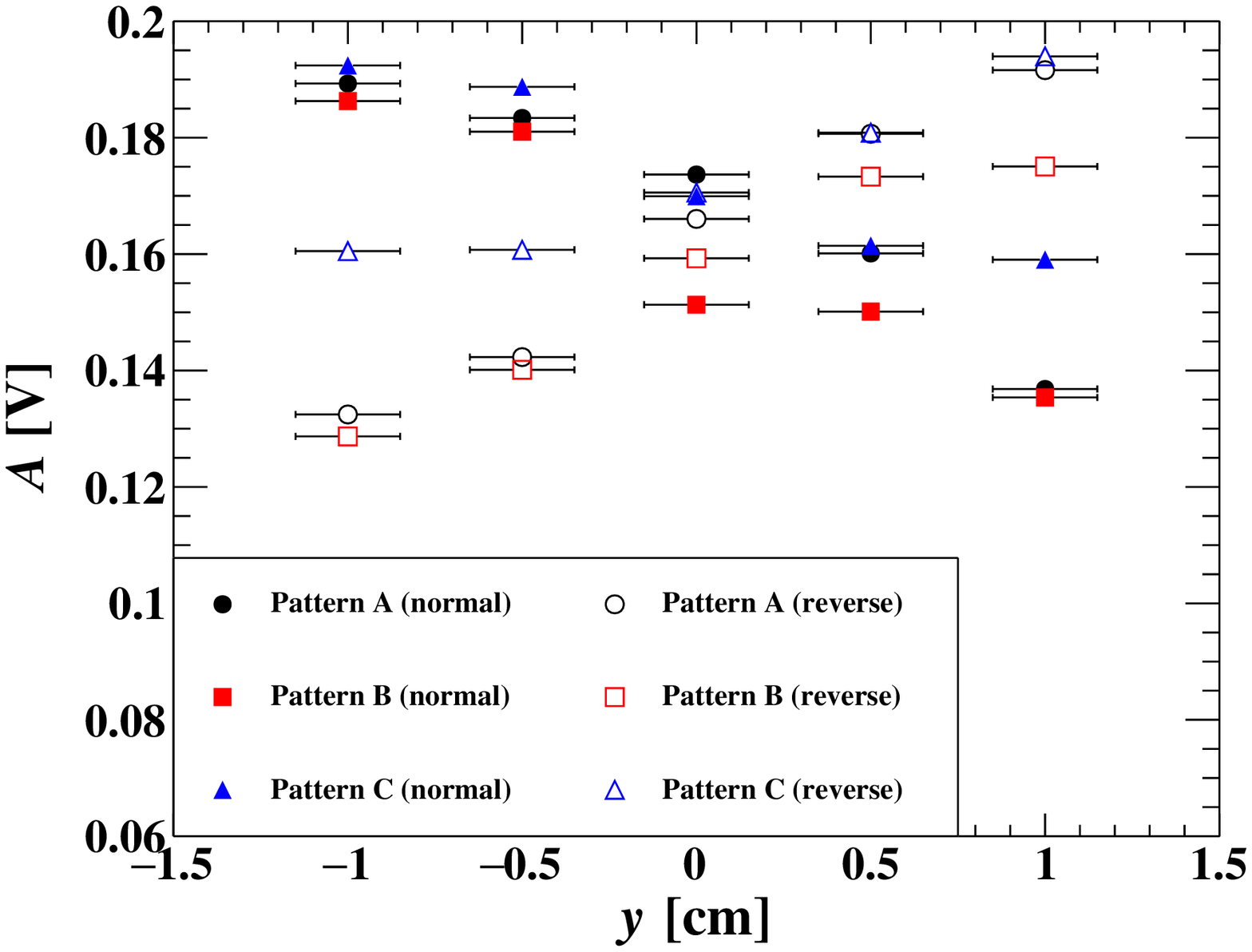}
	\caption{Position dependence of the pulse height in a counter at the nearest point to the SiPMs 
        ($x = -4.25$~cm). The applied voltage was fixed to 162.5~V at 30$^\circ$C.}
	\label{PositionHeightNear}
\end{minipage} 
\hfill
\begin{minipage}[t]{\columnwidth}
	\centering
	\includegraphics[width=\columnwidth]{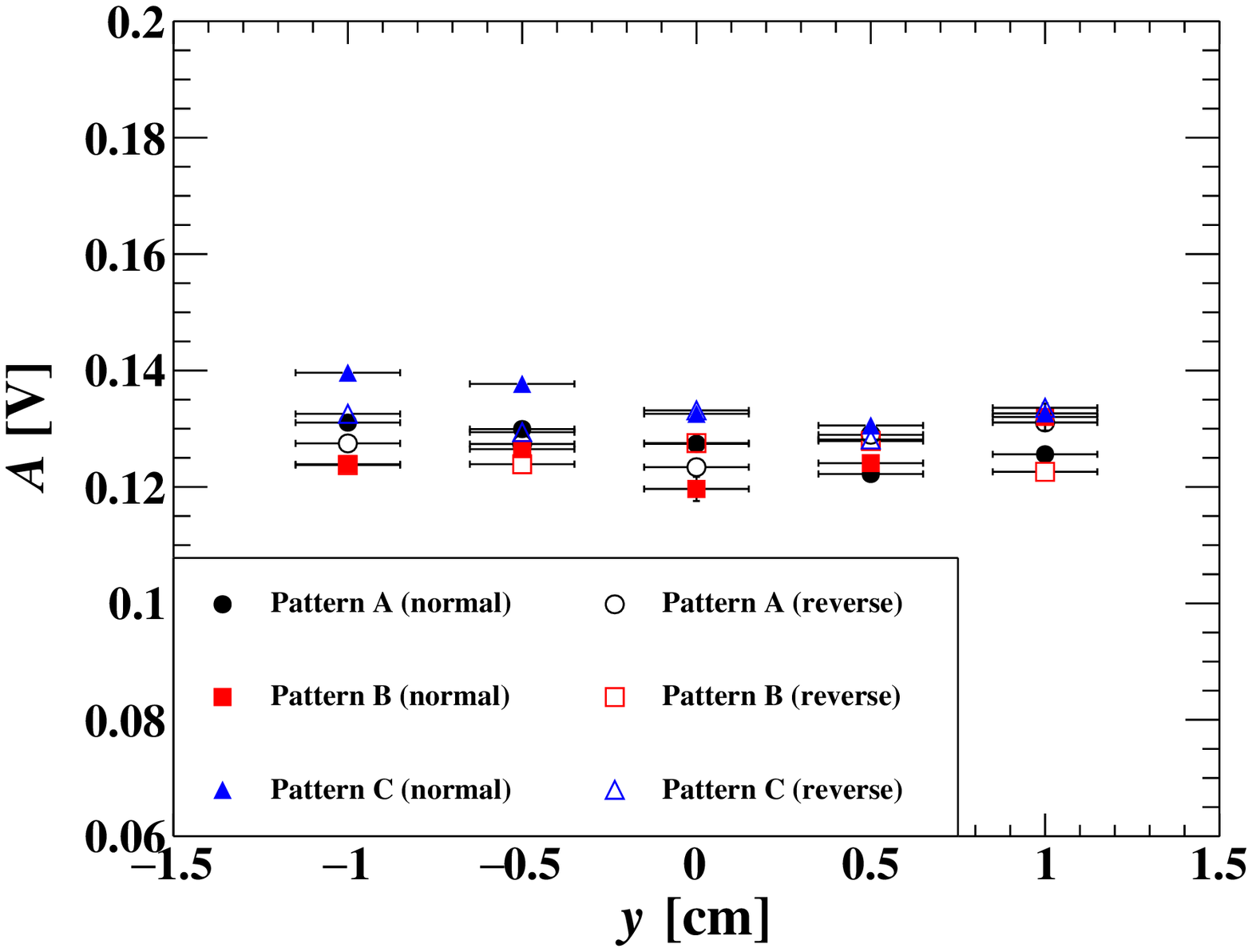}
	\caption{Position dependence of the pulse height in a counter at $x = 0.00$~cm. 
        The applied voltage was fixed to 162.5~V at 30$^\circ$C.}
	\label{PulseHeightCenter}
\end{minipage} 
\end{figure*}

\Fref{PositionHeightNear} shows the dependence of the pulse height on $y$-position 
at $x = -4.25~\mathrm{cm}$. 
A clear dependence is observed, and the slope is opposite for the normal and the reverse order cases, clearly indicating the effect of the position-dependent damage to SiPMs.
When the illuminated point is close to more-damaged SiPMs 
the pulse height becomes smaller since the damaged SiPMs, which detect a large fraction of scintillation light in this case, have smaller gain due to the smaller over-voltages. 

When the illuminated points move far from the SiPMs, the $y$ dependence of the pulse height decreases
and at $x=0.00$~cm, the dependence becomes almost flat as shown in \fref{PulseHeightCenter}.

\subsubsection*{Time centre}
\label{sec:time-center}
\begin{figure*}[tb]
\begin{minipage}[t]{\columnwidth}
	\centering
	\includegraphics[width=1.\columnwidth]{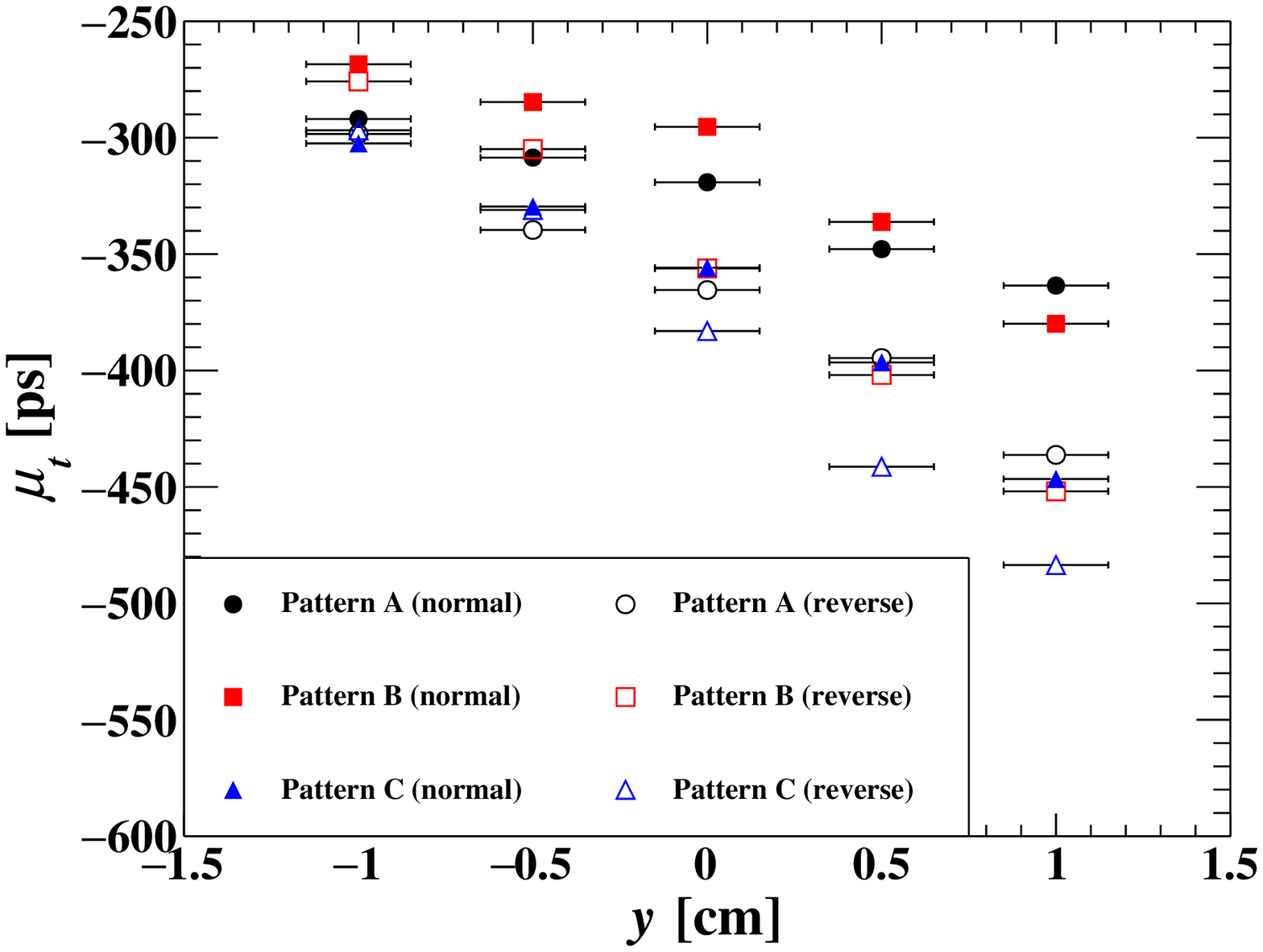}
	\caption{Position dependence of the time centre in a counter at $x = -4.25$~cm. 
        The applied voltage was fixed to 162.5~V at 30$^\circ$C.}
	\label{TimeCenter2cm}
\end{minipage} 	
\hfill
\begin{minipage}[t]{\columnwidth}
	\centering
	\includegraphics[width=1.\columnwidth]{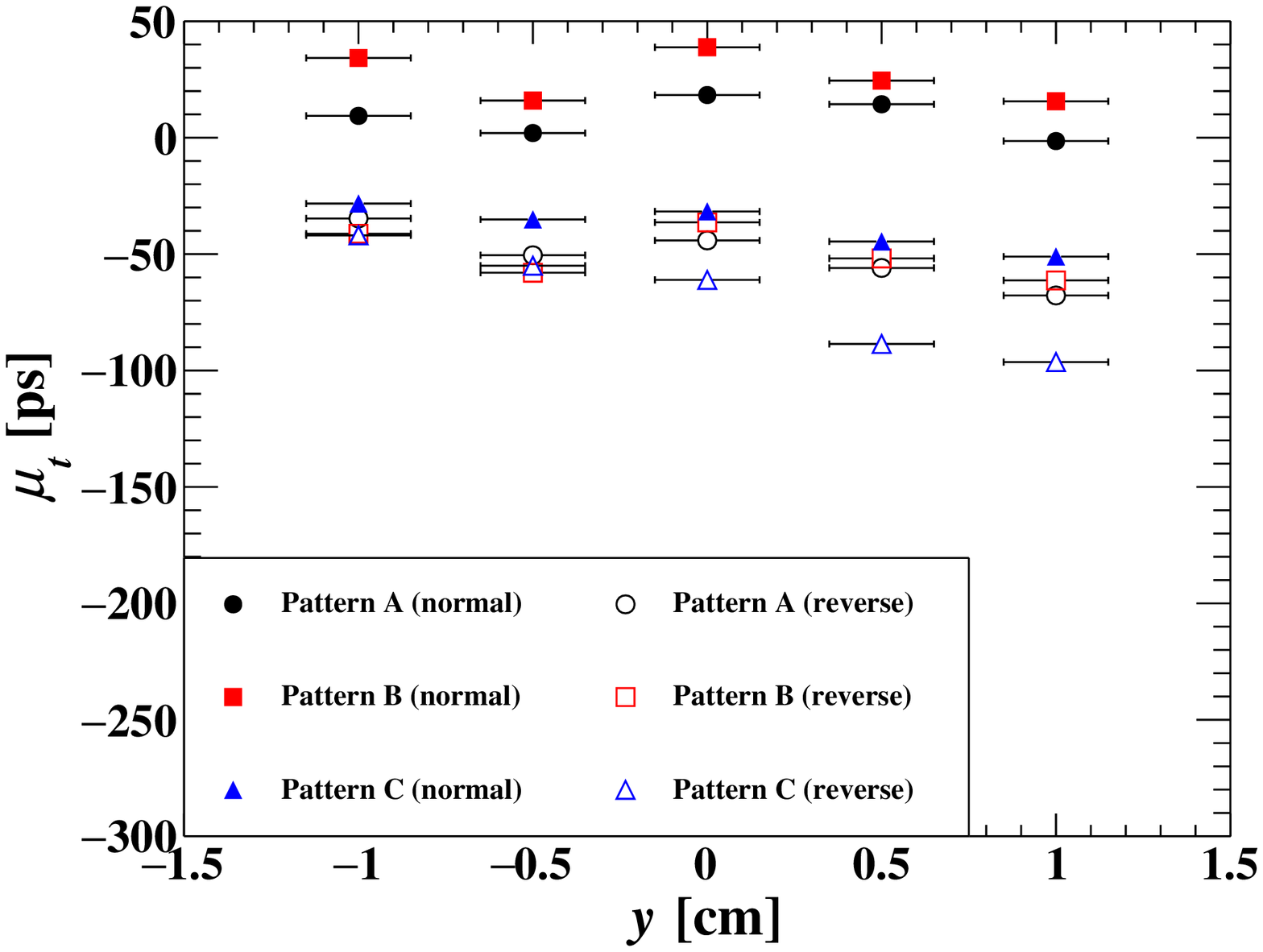}
	\caption{Position dependence of the time centre in a counter at $x = 0.00$~cm. 
        The applied voltage was fixed to 162.5~V at 30$^\circ$C.}
	\label{TimeCenter0cm}
\end{minipage} 	
	
\end{figure*}

\Frefsa{TimeCenter2cm}{TimeCenter0cm} show the time centre $\mu_t$ at $x=-4.25$~cm and $x=0.00$~cm, 
respectively. 
Similar $y$-dependencies are observed in all the damage patterns at $x =-4.25$~cm. 
For this variable, however, the results of normal- and reverse-order show the same sign of the slope, while the magnitudes of the slope are different.
The mechanism yielding the dependence is more complex than that for pulse height
because the signal timing in the CFD method is determined by multiple factors: the distribution and arrival time of photons to each SiPM, the gain of each SiPM, the time response of each SiPM, and the signal propagation time along the series-connected SiPMs.

The common trend stems from a finite propagation time of the scintillation light in the scintillator and the electronic signal through the series-connected SiPMs.
Because of the read out from $+y$ side, signals from higher $y$ hits are detected earlier than those from lower $y$ hits.
This interpretation is confirmed by a measurement using non-irradiated SiPMs.

\begin{figure}[tbp]
	\centering
	\includegraphics[width=1\columnwidth]{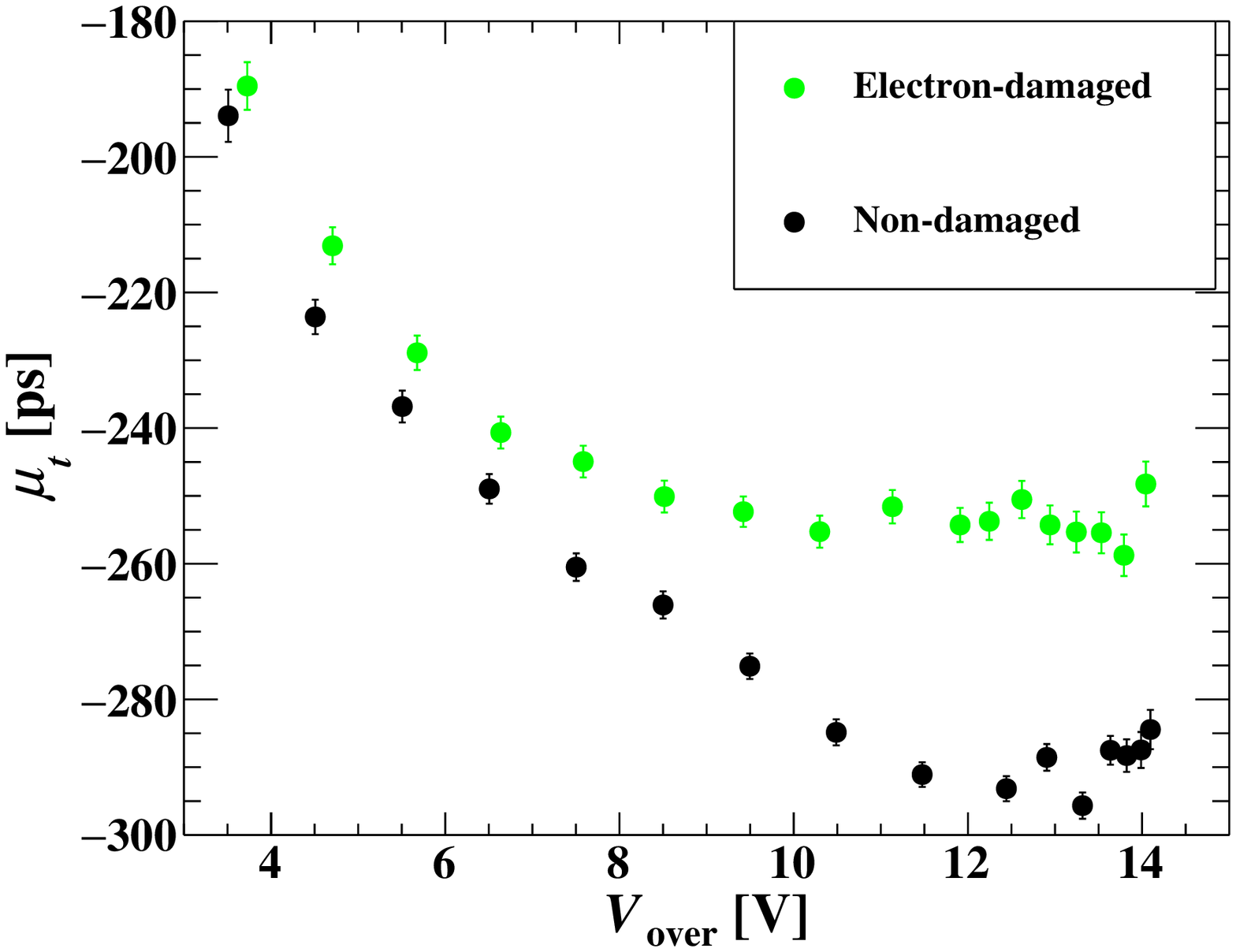}
	\caption{Time centre versus over-voltage of four series-connected SiPMs at 30$^\circ$C. The non-damaged SiPMs are \#11--\#14, while the damaged ones are \#1, \#3--\#5.}
	\label{TimeCenterScanSingle}
\end{figure}

Deviations from the common trend suggest the effect of radiation damage but the non-uniform 
radiation damage complicates matters because of the over-voltage variation among the SiPMs, affecting both the gain and the time response.
To measure the relationship between the over-voltage and the time response, the over-voltage dependence of the time centre for the all-non-damaged case and the all-damaged case is measured. Because of the limited number of non-damaged samples, this measurement was performed using four SiPMs in series. The result is shown in \fref{TimeCenterScanSingle}.
Both damaged and non-damaged SiPMs show a faster time response with higher over-voltage, although it saturates at some point. This suggests faster buildup and propagation of avalanche in a stronger electric field \cite{avalancheDiffuse, avalancheModel}.
Due to the dependence of $V_\mathrm{over}$ on $\mu_t$, SiPMs with less damage in a chain output faster signals than the ones with more damage. 

A qualitative explanation for the difference between the normal and the reverse order is the following.
The measurement of the timing at the 20\% CFD fraction samples the leading edge of the pulse. 
The main contribution to this part comes from few SiPMs closest to the signal readout line 
(i.e.\ $+y$ side), especially when the hit point is at higher $y$.
In the normal order case, these SiPMs yield smaller and slower signals and thus the time centre 
is larger than that in the reverse order case. 
When the hit point is at lower $y$, a smaller fraction of the scintillation light is collected by 
these SiPMs and signals from successive SiPMs contribute more to the CFD timing.
This weakens the effect of the position-dependent damage. Therefore, similar time centre values are obtained 
in the two cases.
As a general trend, the normal order yields less position dependence than the reverse order since 
the position-dependent damage effect compensates the common trend.
The position dependence is mitigated when the fraction of CFD is increased.

When the illuminated point moves far from the SiPMs, 
the position dependence in $y$-direction becomes gradually decreases. 


\subsubsection*{Timing resolution}
\begin{figure*}[tbp]
\begin{minipage}[t]{\columnwidth}
	\centering
	\includegraphics[width=\columnwidth]{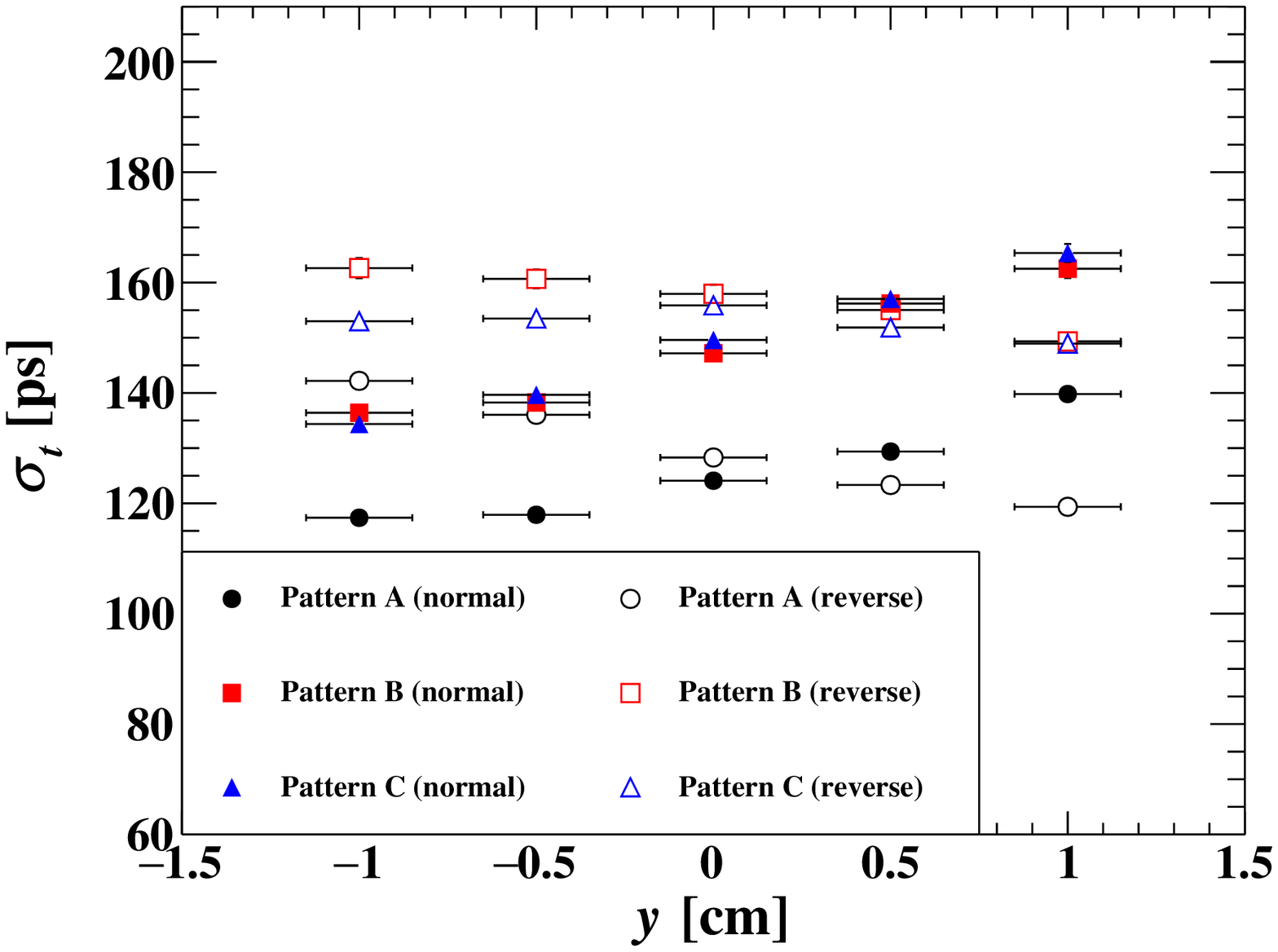}
	\caption{Position dependence of the timing resolution in a counter at $x = -4.25$~cm. 
        The applied voltage was fixed to 162.5~V at 30$^\circ$C.}
	\label{TimeResolutionNear}
\end{minipage} 
\hfill
\begin{minipage}[t]{\columnwidth}
	\centering
	\includegraphics[width=\columnwidth]{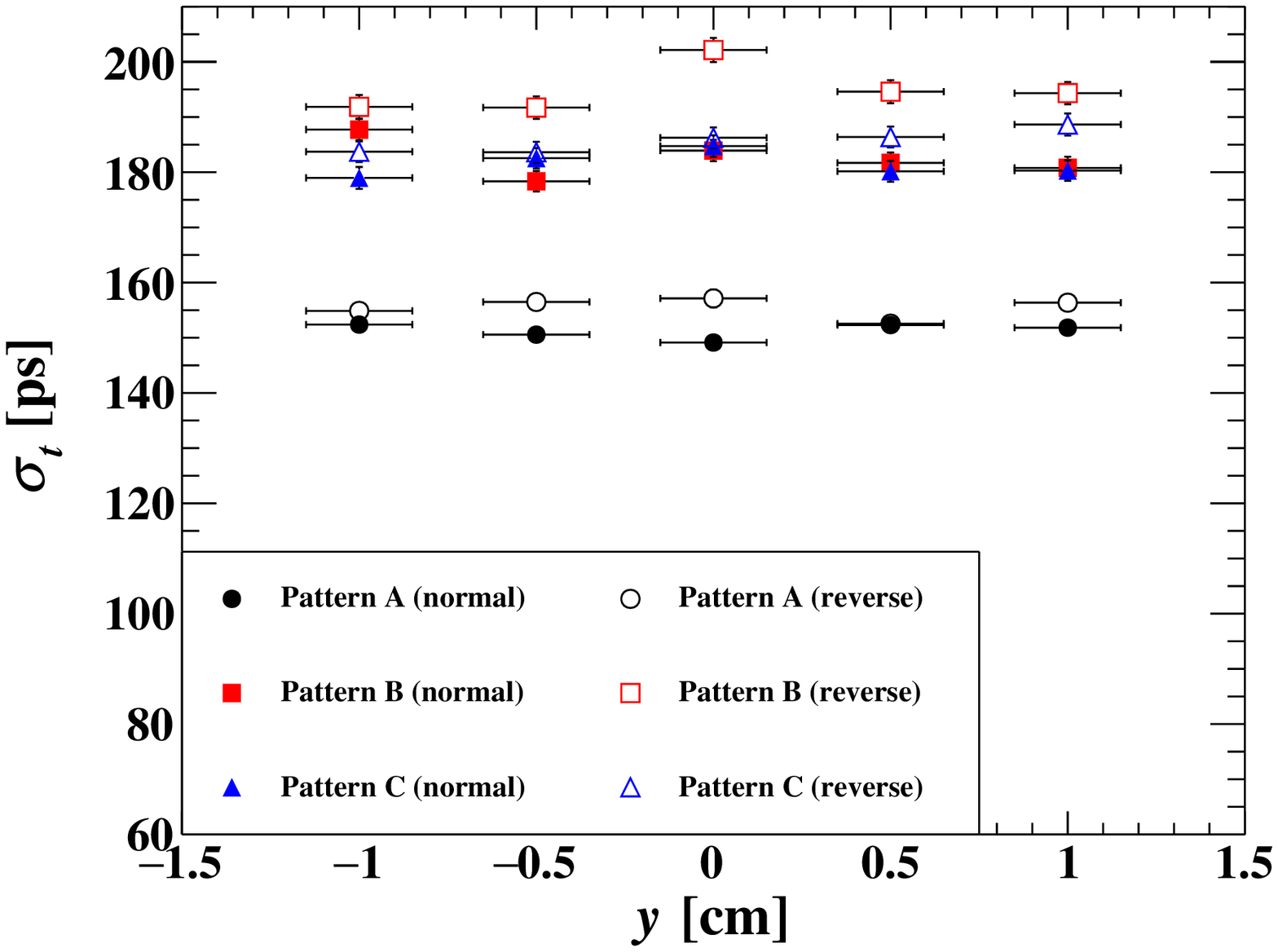}
	\caption{Position dependence of the timing resolution in a counter at $x = 0.00$~cm. 
        The applied voltage was fixed to 162.5~V at 30$^\circ$C.}
	\label{TimeResolutionCenter}
\end{minipage} 
\end{figure*}

\Frefsa{TimeResolutionNear}{TimeResolutionCenter} show the position dependence of the timing resolution
at $x = -4.25$ and $0.00~\mathrm{cm}$, respectively. 
The flipped behaviour between the normal and the reverse order cases is due to 
the timing resolution predominantly depending on the signal-to-noise ratio. It becomes worse when the illuminated point is close to the damaged SiPMs. 

The data show slightly better resolutions in the normal order than in the reverse order. 
This may indicate a second-order effect on the timing resolution from the combination of SiPMs with different time responses.
As discussed in Sect.~\ref{sec:time-center}, the normal order connection compensates the dispersion of signal timings measured by different SiPMs, resulting in a better resolution.


\begin{figure}[tbp]
	\centering
	\includegraphics[width=\columnwidth]{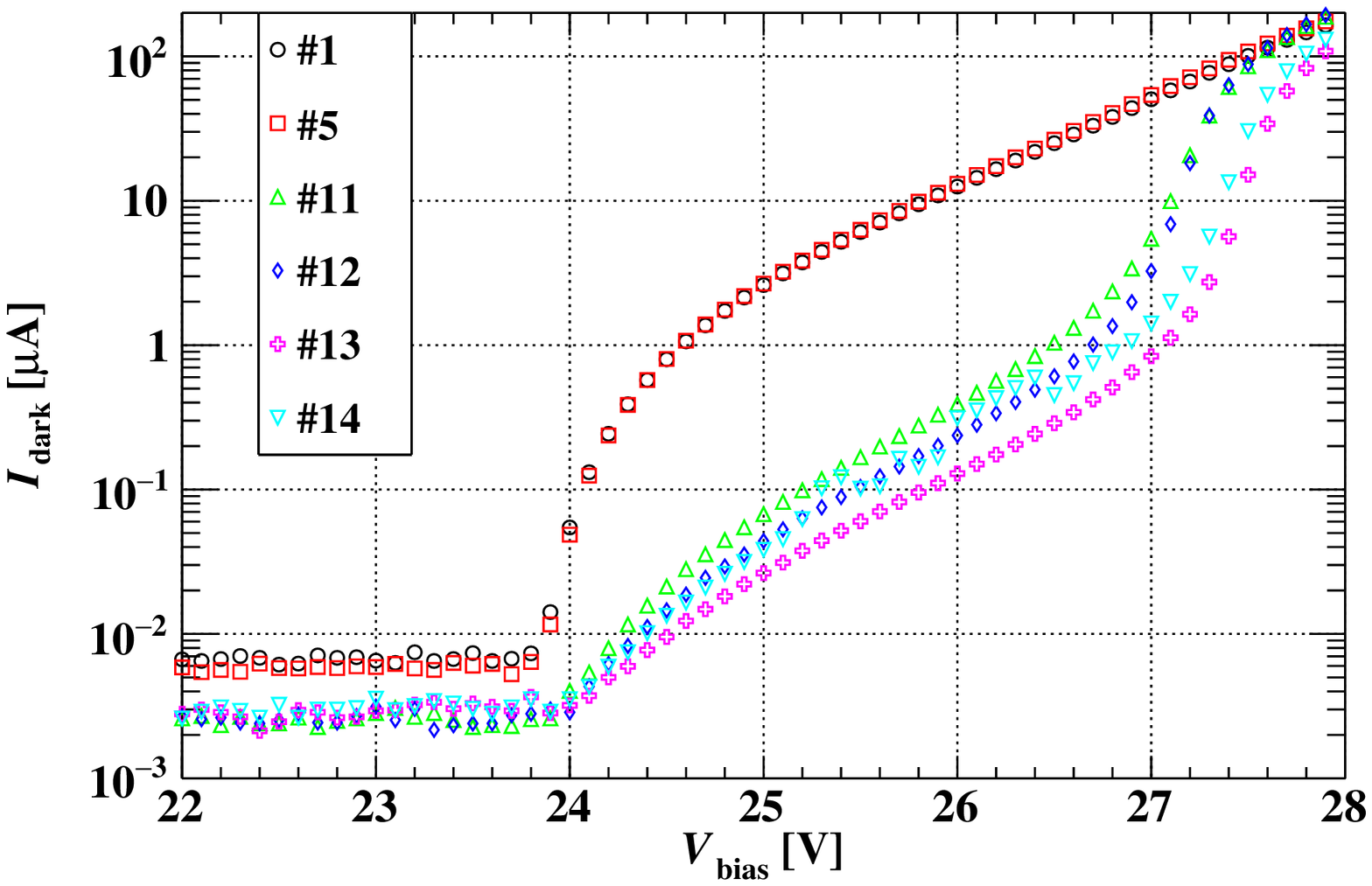}
	\caption{I-V curves of each SiPM used in pattern A at 10$^\circ$C.}
	\label{IV_patternA_10deg}
\end{figure}

\section{Discussion}

The timing resolution $\sigma_t$ is predominantly determined by the signal-to-noise ratio.
Irradiation induces an increase in $I_\mathrm{dark}$ and causes deterioration of $\sigma_t$; the deterioration is described by the square root of $I_\mathrm{dark}$ dependence. 
This indicates that in a running experiment $\sigma_t$ can be estimated
in real time by monitoring $I_\mathrm{dark}$.

As expected, cooling helps to reduce the impact of the radiation damage on $\sigma_t$.
Because of limitations in the setup, the only tested temperature was 10$^\circ$C.
Cooling down to lower temperatures further reduces the impact or enables operation in a harsher environment, although 
it imposes additional constraints, such as thermal insulation and dehumidification, on the detector design.
In addition, re-optimisation of $V_\mathrm{bias}$ and the CFD fraction values depending on the damage level
 turns out to be effective in mitigating the deterioration of $\sigma_t$.

Annealing of SiPMs can be another way to recover the original $\sigma_t$. However, if the SiPMs are glued to the scintillator with optical cement,
the applicable temperature is limited to well below 70$^\circ$C, at which the scintillator melts.

A study on connecting differently-damaged SiPMs in series was performed in \cite{Mu2e_Experiment},
which reports the measurement of the charge for laser light readout by three SiPMs connected in series: two of them were non-damaged and one was extensively damaged.
The present study extends it systematically and quantitatively to find additional effects on the operation and performance of scintillator counters. 

An apparent breakdown-voltage shift is found, which is a phenomenon peculiar to the series connection of differently-damaged SiPMs, 
spoiling the determination of the effective breakdown voltage from the I-V characteristics.
Monitoring the operating over-voltage is important for long-term data taking. Therefore, this phenomenon should be monitored during the operation of detectors in a real experiment.

As a whole, counters equipped with differently-damaged SiPMs work as properly as those with equally-irradiated SiPMs. 
The pulse height does not change on average and the timing resolution is similar to that of equally-irradiated SiPMs with the equivalent damage level. 
These can be an advantage of the series connection over parallel connection. 
In parallel connection, each SiPM has a fixed over-voltage, leading to the same pulse height,
however, the currents $I_\mathrm{dark}$ are highly non-uniform among the SiPMs, and the operating voltage and the timing resolution can be limited.

A drawback of the series connection is the position dependence of the pulse height, the time centre, and the timing resolution, which can be avoided in the case of parallel connection by a careful design of the circuit.
The dependence of the pulse height is peculiar to the case with differently-damaged SiPMs, while that of the time centre stems from a combination of an intrinsic property of series connection and 
effects of the different damage levels.
The intrinsic one observed in this study is in contrast to the result reported in \cite{Series_Connection_Study_MEG}. The difference could be due to the different sizes of the scintillator tile and different numbers of SiPMs in series.

The variation of the time centre in a counter results in a biased time reconstruction depending on the hit position,
and thus, this additional spread in the measured time worsens the timing resolution.
If the hit position information is available, e.g.\ from an external tracker, the dependence can be corrected.
The dependence changes with different damage levels and the combination of SiPMs.
Therefore, the position dependence of the time centre should be monitored during an experiment.
The normal order, i.e.\ extracting signal from the higher damage side, is better because it yields less position dependence than the reverse order. 
Cooling does not help to reduce the position dependence because the difference of over-voltages between the damaged SiPMs and non-damaged SiPMs does not decrease
even when the SiPMs are cooled as shown in \fref{IV_patternA_10deg} and Table~\ref{Tab:Vbr_shift_table}. 

\subsubsection*{Impact on MEG II pTC}

The single-counter $\sigma_t$ deteriorates of 41\% 
by the increase in $I_\mathrm{dark}$ at $V_\mathrm{over} \sim 16.5$~V (30$^\circ$C) after irradiation of $\Phi_\mathrm{eq}\approx 3 \times 10^{9}~\mathrm{cm^{-2}}$, which is equivalent to the average damage level in the MEG II three-year run. This deterioration is reduced to 13\% by cooling the setup to 10$^\circ$C.
The pTC system is equipped with a water chiller system, which can control the detector 
temperature in the range $8$ -- $25^\circ$C while it is expected to be $\sim 30^\circ$C without it. 
The MEG II collaboration adopted 10$^\circ$C as the operating temperature of the pTC.

The closest approximation to the damage in the pTC channels is pattern C since the positron flux is larger at smaller 
radius. 
Unfortunately, the pTC was designed with the readout from the larger radius side, which corresponds to the reverse order case in this study, and all the counters have been already assembled.
A $y$-dependence of $\sim 80~\mathrm{ps/cm}$ is expected at $x=-4.25$~cm.
The external tracking detector will be available to estimate the 
hit $y$-position. Therefore, a correction for the time centre dependence 
can be applied and the deterioration can be, at least partially, compensated.

\section{Conclusions}

The effect of radiation damage to SiPMs on the performances of a
scintillator counter with series-connected readout was studied
using ASD-NUV3S-P High Gain SiPMs irradiated either with electrons from $^{90}$Sr sources 
up to  $\Phi_\mathrm{e^-}\approx 3 \times 10^{12}~\mathrm{cm}^{-2}$ or 
with neutrons from a reactor up to $\Phi_\mathrm{eq}\approx 5.5 \times 10^9~\mathrm{cm^{-2}}$.
The increase in dark count rate is the main cause of the timing resolution deterioration. 
Reducing the dark count rate by cooling the SiPMs effectively recovers the timing resolution. 
Position dependencies of the pulse height, the time centre and the timing resolution are 
observed when the series-connected SiPMs have different I-V characteristics. 
The time centre dependence on the hit position results in additional degradation of the timing resolution.
When the damage level is not uniform within the series-connected SiPMs, the time centre depends 
also on the direction of the signal line.
The best timing resolution is obtained by reading the signal from the side where the most heavily
damaged SiPMs are located. 
Furthermore, a shift of the apparent breakdown voltage is observed, which might spoil the 
determination of the effective breakdown voltage from the I-V curve.

\section*{Acknowledgments}
We thank PSI as the host laboratory, and we thank Arnold Marcel from the PSI group who helped us to 
calculate the source activities. 
We thank the Laboratorio Energia Nucleare Applicata (LENA) in Pavia
and in particular Alloni Daniele, Prata Michele, and Salvini Andrea for the irradiation of SiPMs with neutrons.
This study was supported by INFN, JSPS KAKENHI Grant Numbers JP26000004, JP19J13635, 
and JSPS Core-to-Core Program, A. Advanced Research Networks JPJSCCA20180004.



\begin{thebibliography}{99}
\expandafter\ifx\csname url\endcsname\relax
  \def\url#1{\texttt{#1}}\fi
\expandafter\ifx\csname urlprefix\endcsname\relax\def\urlprefix{URL }\fi

\bibitem{Series_Connection_Study_MEG}
P.\,W. Cattaneo et~al., Development of high precision timing counter based on plastic scintillator with SiPM readout, IEEE Trans. Nucl. Sci. 61 (2014) 2657--2666, \url{https://doi.org/10.1109/TNS.2014.2347576}.
\bibitem{MEG_II_Experiment}
A.\,M.~Baldini et~al., The design of the MEG II experiment, Eur. Phys. J. C (2018) 78:380, \url{https://doi.org/10.1140/epjc/s10052-018-5845-6}.
\bibitem{Mu2e_Experiment}
N.~Atanov et al., The Mu2e calorimeter final technical design report, arXiv:1802.06341
\bibitem{PANDA_TOF_Detector}
M.~B\"ohm et~al., Fast SiPM readout of the PANDA TOF detector, 2016 J. Instrum. 11 C05018, \url{https://doi.org/10.1088/1748-0221/11/05/c05018}.
\bibitem{Radiation_Damage_Recent_Paper}
 E.~Garutti, Yu.~Musienko, Radiation damage of SiPMs, Nucl. Instrum. Methods A 926 (2019) 69--84, \url{https://doi.org/10.1016/j.nima.2018.10.191}.

 
\bibitem{pilotrun2016}
 M.~Nakao et~al., Results from pilot run for MEG II positron timing counter, Springer Proc. Phys. 213 (2018) 237--241, \url{https://doi.org/10.1007/978-981-13-1316-5}.
\bibitem{EffectiveNIEL}
C.~Inguimbert, P.~Arnolda, T.~Nuns, G.~Rolland, ``Effective NIEL'' in silicon: calculation using molecular dynamics simulation results, IEEE Trans. Nucl. Sci. 57 (2010) 1915--1923, \url{https://doi.org/10.1109/TNS.2010.2049581}.
\bibitem{Pisa_Proceedings}
M.~Usami et al., Radiation damage effect on timing resolution of 6 series-connected SiPMs for MEG II positron Timing Counter, Nucl. Instrum. Methods A 936 (2019) 572--573, \url{https://doi.org/10.1016/j.nima.2018.10.053}.
\bibitem{VCI2016}
Y.~Uchiyama et al., 30-ps time resolution with segmented scintillation counter for MEG II, Nucl. Instrum. Methods A 845 (2017) 507--510, \url{https://doi.org/10.1016/j.nima.2016.06.072}.

\bibitem{ASTM-E-722-93}
ASTM Standard E722-93, Standard practice for characterizing the neutron energy fluence spectra in terms of an equivalent monoenergentic neutron fluence for radiation-hardness testing of electronics, in:
 1993 Annual Book of ASTM Standards, Vol. 12.02, ASTM International, West Conshohocken, PA, 1993, pp. 324--337. 

\bibitem{DRS}
S.~Ritt, R.~Dinapoli, U.~Hartmann, Application of the DRS chip for fast waveform digitizing, Nucl. Instrum. Methods A 623 (2010) 486--488, \url{https://doi.org/10.1016/j.nima.2010.03.045}.

\bibitem{simonetta}
M.~Simonetta et~al., Test and characterisation of SiPMs for the MEGII high resolution Timing Counter, Nucl. Instrum. Methods A 824 (2016) 145--147, \url{https://doi.org/10.1016/j.nima.2015.11.023}. 

\bibitem{avalancheDiffuse}
A.~Lacaita, M.~Mastrapasqua, M.~Ghioni, S.~Vanoli, Observation of avalanche propagation by multiplication assisted diffusion in $p$-$n$ junctions, Appl. Phys. Lett. 57 (1990) 489--491, \url{https://doi.org/10.1063/1.103629}.
\bibitem{avalancheModel}
A.~Spinelli, A.\,L.~Lacaita, Physics and numerical simulation of single photon avalanche diodes, IEEE Trans. Electron Devices 44 (1997) 1931--1943, \url{https://doi.org/10.1109/16.641363}.
\end{thebibliography}
\end{document}